\documentclass[preprintnumbers,article,amsmath,amssymb,floatfix,10pt,prd,twocolumn,superscriptaddress,nofootinbib,times,aps,eqsecnum]{revtex4-2}
\usepackage{subcaption}
\usepackage{amsfonts}
\usepackage{latexsym}
\usepackage[utf8]{inputenc}
\usepackage{graphicx}
\usepackage{amsmath}
\usepackage{palatino}
\usepackage{mathpazo}
\usepackage{textcomp}
\usepackage{float}
\usepackage{booktabs}
\usepackage{dcolumn}
\usepackage{ragged2e}
\usepackage{hyperref}
\hypersetup{colorlinks,citecolor=blue}
\hypersetup{colorlinks=true,linkcolor=blue,filecolor=magenta,urlcolor=blue}
\usepackage{amsmath}

\usepackage[]{natbib}
\usepackage{xcolor}
\usepackage{orcidlink}
\usepackage{caption}
\usepackage{subcaption}

\usepackage[toc]{appendix}
\usepackage{commath}
\usepackage{cancel}
\usepackage{csquotes}
\usepackage{placeins}
\usepackage{multirow}

\def\jnl@style{\it}
\def\aaref@jnl#1{{\jnl@style#1}}

\def\aaref@jnl#1{{\jnl@style#1}}

\def\aj{\aaref@jnl{AJ}}                   
\def\apj{\aaref@jnl{ApJ}}                 
\def\apjl{\aaref@jnl{ApJ}}                
\def\apjs{\aaref@jnl{ApJS}}               
\def\apss{\aaref@jnl{Ap\&SS}}             
\def\aap{\aaref@jnl{A\&A}}                
\def\aapr{\aaref@jnl{A\&A~Rev.}}          
\def\aaps{\aaref@jnl{A\&AS}}              
\def\mnras{\aaref@jnl{Mon.~Not.~Roy.~Astron.~Soc.}}             
\def\prd{\aaref@jnl{Phys.~Rev.~D}}        
\def\prc{\aaref@jnl{Phys.~Rev.~C}}  
\def\prl{\aaref@jnl{Phys.~Rev.~Lett.}}    
\def\qjras{\aaref@jnl{QJRAS}}             
\def\skytel{\aaref@jnl{S\&T}}             
\def\ssr{\aaref@jnl{Space~Sci.~Rev.}}     
\def\zap{\aaref@jnl{ZAp}}                 
\def\nat{\aaref@jnl{Nature}}              
\def\aplett{\aaref@jnl{Astrophys.~Lett.}} 
\def\apspr{\aaref@jnl{Astrophys.~Space~Phys.~Res.}} 
\def\physrep{\aaref@jnl{Phys.~Rep.}}      
\def\physscr{\aaref@jnl{Phys.~Scr}}       
\def\commat{\aaref@jnl{Comm.~Math.~Phys.}}              
\def\science{\aaref@jnl{Science}}               
\def\cqg{\aaref@jnl{Classical Quant.~Grav.}}            
\def\jpcs{\aaref@jnl{JPCS}}                                     
\def\ijmpd{\aaref@jnl{Int.~J.~Mod.~Phys.~D}}                    
\def\grg{\aaref@jnl{Gen.~Relat.~Gravit.}}               
\def\rpp{\aaref@jnl{Rep.~Prog.~Phys.}}          
\def\npa{\aaref@jnl{Nucl.~Phys.~A}}        
\def\lrr{\aaref@jnl{Living Rev.~Rel.}}                   
\def\jcap{\aaref@jnl{J.~Cosmology Astropart.~Phys.}}    
\def\rmp{\aaref@jnl{Rev.~Mod.~Phys.}}   
\def\epjc{\aaref@jnl{Eur.~Phys.~J.~C}}
\def\plb{\aaref@jnl{~Phy.~Lett.~B}}
\def\mpla{\aaref@jnl{Mod.~Phy.~Lett.~A}}
\def\arxiv{\aaref@jnl{arxiv.org}}

\allowdisplaybreaks[1]

\begin{document}

\color{blue}

\title{$f(Q,\mathcal{L}_m)$ gravity, and its cosmological implications}

\author{Ayush Hazarika\orcidlink{0009-0004-5255-0730}}
\email{ayush.hazarika4work@gmail.com}
\affiliation{Department of Physics, Tezpur University, Napaam, Tezpur, 784028, Assam, India}

\author{Simran Arora\orcidlink{0000-0003-0326-8945}}
\email{dawrasimran27@gmail.com}
\affiliation{Center for Gravitational Physics and Quantum Information, Yukawa Institute for Theoretical Physics,\\ Kyoto University, 606-8502, Kyoto, Japan.}

\author{P.K. Sahoo\orcidlink{0000-0003-2130-8832}}
\email{pksahoo@hyderabad.bits-pilani.ac.in}
\affiliation{Birla Institute of Technology and Science, Pilani, Hyderabad Campus, Jawahar Nagar, Kapra Mandal, Medchal District, Telangana 500078, India.}

\author{Tiberiu Harko\orcidlink{0000-0002-1990-9172}}
\email{tiberiu.harko@aira.astro.ro}
\affiliation{Department of Physics, Babes-Bolyai University, 1  Kogalniceanu Street, 400084 Cluj-Napoca, Romania}
\affiliation{Astronomical Observatory, 19 Ciresilor Street, 400487, Cluj-Napoca, Romania}

\date{\today}

\begin{abstract}
Symmetric teleparallel gravity and its $f(Q)$ extensions have emerged as promising alternatives to General Relativity (GR), yet the role of explicit geometry–matter couplings remains largely unexplored. In this work, we address this gap by proposing a generalized $f(Q,\mathcal{L}_m)$ theory, where the gravitational Lagrangian density depends on both the non-metricity scalar $Q$ and the matter Lagrangian $\mathcal{L}_m$. This formulation naturally includes Coincident GR and the Symmetric Teleparallel Equivalent of GR as special cases. Working in the metric formalism, we derive the corresponding field equations, which generalize those of the standard $f(Q)$ gravity, and obtain the modified Klein–Gordon equation for scenarios involving scalar fields. The cosmological implications of the theory are explored in the context of the Friedmann–Lemaître–Robertson–Walker (FLRW) universe. As a first step, we obtain the modified Friedmann equations for $f(Q,\mathcal{L}_m)$ gravity in full generality. We then investigate specific cosmological models arising from both linear and non-linear choices of $f(Q,\mathcal{L}_m)$, performing detailed comparisons with the standard $\Lambda$CDM scenario and examining their observational consequences.
\end{abstract}

\maketitle

\tableofcontents

\section{Introduction}\label{sec:I}

The theory of General Relativity (GR) \cite{einstein1915feldgleichungen, hilbert1915grundlagen, einstein1916foundation} revolutionized our understanding of gravity by conceptualizing it not as a conventional force but as an inherent property of spacetime, rooted in Riemannian geometry \cite{wald2010general}. Thus, GR developed, contrary to Einstein's initial intent of not ``geometrizing gravity" \cite{lehmkuhl2014einstein}. In GR, the metric and matter interact minimally, as defined by the Einstein-Hilbert action ($S_{EH}$), given as $S_{EH}= (1/2\kappa) \int \sqrt{-g}R\,d^4x + S_m$, where $\kappa$ denotes the term of the gravitational coupling constant. $g$ is the determinant of the metric tensor and $R$ represents the Ricci scalar. By $S_m$, we have denoted the action of the matter. This linearity in $R$ gives a second-order field equation $G_{\mu \nu}=R_{\mu\nu}-(1/2)R\,g_{\mu\nu}=\kappa T_{\mu\nu}$, which governs the dynamics of matter in the curved spacetime and relates geometry, described by the Einstein tensor $G_{\mu \nu}$ to the matter energy-momentum tensor $T_{\mu \nu}$.
\par A large number of observations have established GR as a very successful theory of gravity by confirming many of its predictions, such as the deflection of light by the Sun's gravitational field \cite{light_deflection}, the perihelion motion of Mercury \cite{2021arXiv211111238J}, the existence of gravitational waves \cite{abbott2016ligo}, gravitational redshift \cite{landau2013classical}, orbital decay of the Hulse-Taylor binary pulsar \cite{taylor1982new}, and the radar echo delay \cite{shapiro1971general,shapiro1976verification}, respectively. An in-depth analysis of all the experimental and observational tests of GR can be found in \cite{will2006living,ishak2019testing}.

Despite its remarkable success, which spanned almost one hundred years, the theory of GR currently faces numerous challenges. At a quantum level, it cannot explain the quantum properties of the gravitational interaction \cite{rovelli2004quantum}. Also, gravitational collapse can result in geodesic incompleteness under specific assumptions regarding the energy-momentum tensor \cite{hawking1970singularities}. This implies that certain types of geodesics are constrained by an upper limit on an affine parameter, indicating a singular structure in spacetime. One notable consequence of this phenomenon is the appearance of cosmological singularities during the Big Bang \cite{borde1994eternal} and the existence of black holes \cite{penrose1969gravitational}.\par
A significant challenge for GR did appear when it was faced with the problem of explaining the late-time cosmic accelerated expansion. Evidence from observations of type Ia supernovae \cite{riess1998observational,perlmutter2003measuring,tonry2003cosmological}, large-scale structure observations, and measurements of the cosmic microwave background (CMB) anisotropies from the Wilkinson Microwave Anisotropy Probe (WMAP) \cite{spergel2007three}, and of the Planck satellite \cite{aghanim2020planck} highlighted a limitation in GR's ability to fully describe and comprehend the dynamics of the Universe at cosmic scales during its later stages. This failure of GR prompted the exploration of alternative theories of gravity and the consideration of additional factors or components in the gravitational field equations, such as dark energy or the addition of a cosmological constant ($\Lambda$) in action ($S_{EH}$), to reconcile observations with theoretical predictions \cite{wald2010general}. Hence, a more general gravitational framework is required to explain the gravitational dynamics across various scales, ranging from the Solar System to galaxies and the large scale Universe.\par
In pursuit of a more comprehensive understanding of gravity that aligns with the observational evidence, a plethora of modified gravity theories have been proposed, such as Scalar-tensor theories \cite{faraoni2004scalar,brans1961mach,dicke1962mach,bergmann1968comments,nordtvedt1970post,wagoner1970scalar}, Tensor-Vector-Scalar (TeVeS) \cite{bekenstein2004relativistic}, Dvali-Gabadadze-Porrati (DGP) gravity \cite{dvali20004d}, Einstein-Gauss-Bonnet gravity \cite{glavan2020einstein}, brane-world gravity \cite{maartens2010brane}, Einstein-Aether theory \cite{jacobson2001gravity}, Eddington-inspired Born-Infeld (EiBI) gravity \cite{banados2010eddington,jimenez2018born}  etc.
\par A class of gravitational theories, known as $f(R)$ gravity, arises through a straightforward extension of the Einsetin-Hilbert action $S_{EH}$ by replacing $R$ with an arbitrary functions of the Ricci scalar $R$ \cite{sotiriou2010f}. The geometrical structures of the $f(R)$ gravity were able to explain the accelerated cosmic expansion \cite{song2007large},  and also the flat rotation curves of galaxies, without introducing dark matter \cite{bohmer2008dark}. Even though it failed when subjected to Solar-System tests \cite{berry2011linearized,chiba2007solar,guo2014solar}, $f(R)$ gravity could still be a valuable approach to the foundational framework for a ``parameterized post-Friedmann" description of linear phenomena and could draw parallels with the parameterized post-Newtonian framework for small-scale tests of gravity.
\par An alternative method for extending the Einstein-Hilbert action involves postulating the presence of a non-minimal coupling between geometry and matter, and it leads to the $f(R,\mathcal{L}_m)$ gravity \cite{harko2010f}. For the various astrophysical and cosmological implications of this theory see \cite{wang2012energy, jaybhaye2022cosmology, jaybhaye2022constraints, jaybhaye2023baryogenesis, kavya2023static, maurya2023constrained, solanki2023wormhole, singh2023constrained, harko2013extended, maurya2023accelerating}. Another similar approach is based on the inclusion of a non-minimal coupling between geometry, described by the Ricci scalar $R$, and the trace of the energy-momentum tensor $T$, giving rise to $f(R,T)$ gravity \cite{harko2011f}. A more comprehensive exploration of this theory is available in the detailed investigations presented in \cite{myrzakulov2012frw, alvarenga2012testing, zaregonbadi2016dark, fisher2019reexamining, shabani2014cosmological, alvarenga2013dynamics, moraes2017modeling, shabani2017stability, bhattacharjee2020inflation, mahapatra2024neutron}. In all these extended theories, the gravitational dynamics is described by more general functions of the curvature scalar, matter Lagrangian, and the trace of momentum-energy tensor, respectively, which allows for obtaining a broader range of gravitational behaviors going beyond the predictions of GR. For a detailed review of modified gravity and its implications see \cite{carroll2004cosmic, nojiri2007introduction, lobo2008dark, de2010f, sotiriou2010f, capozziello2011extended, olmo2011palatini, nojiri2011unified, bamba2012dark, clifton2012modified, nojiri2017modified, harko2018extensions}.
\par GR is based solely on the metric and on the Riemannian curvature tensor to define gravity. However, within the broader context of metric-affine geometry, gravity is not limited to curvature alone; it can also be mediated by two additional geometric quantities, torsion and non-metricity, respectively.
\par In the context of the Riemannian geometry, the torsion tensor faces a severe limitation. Specifically, due to the symmetry of the Christoffel symbols, the torsion tensor is restricted to zero, that is, $T^{\mu}_{\rho\lambda}=0$. In an interesting extension of Riemann geometry,  in the Weitzenb\"{o}ck space \cite{Weitzenbock:Invariantentheorie}, the torsion tensor is non-zero ($T^{\mu}_{\rho\lambda}\neq 0$), and the Riemann curvature tensor is zero, leading to a spacetime characterized by flat geometry, endowed with a significant property known as absolute parallelism, or teleparallelism. The applications of Weitzenb\"{o}ck-type spacetime in physics were pioneered by Einstein to introduce a unified teleparallel theory, unifying electromagnetism and gravity \cite{einstein1928riemann}. In the teleparallel approach, the fundamental characteristic is the replacement of the metric $g_{\mu\nu}$, which serves as the primary physical variable that describes gravitational properties, with a set of tetrad vectors $e^i_{\mu}$. Torsion, originating from the tetrad fields, can be employed as a comprehensive descriptor of the gravitational effects, thus replacing curvature with torsion. This leads to the theory known as the teleparallel equivalent of general relativity (TEGR) \cite{moller1961conservation, pellegrini1963tetrad, hayashi1979new}, which was extended to the $f(T)$ gravity theory. 
\par In teleparallel or $f(T)$ type gravity theories, torsion exactly compensates for curvature, resulting in a flat spacetime. A notable advantage of $f(T)$ gravity theory lies in its second-order field equations, which differentiates it from $f(R)$ gravity, which, within the metric approach, is described by fourth-order field equations \cite{aldrovandi2012teleparallel}. The applications of $f(T)$ gravity theories have been extensively explored in the study of astrophysical processes and in cosmology. Significantly, these theories are extensively used to provide an alternative explanation for large-scale structure, the late-time accelerating expansion of the Universe, thus eliminating the need to introduce dark energy \cite{li2011large, myrzakulov2011accelerating, cai2016f, ferraro2007modified, ferraro2008born, bengochea2009dark, linder2010einstein, boehmer2012wormhole, harko2014nonminimal, harko2014f, bahamonde2017new, capozziello2017constraining, farrugia2018gravitational, awad2018constant, jimenez2018teleparallel, golovnev2018cosmological, d2018growth, fontanini2019teleparallel, koivisto2019spectrum, pereira2019gauge, blixt2019gauge,https://doi.org/10.1002/prop.202200162}.
\par The third geometric formulation of gravitational theories is based on the non-metricity $Q$ of the metric \cite{nester1998symmetric}. Geometrically, this quantity captures the variation in the length of a vector during parallel transport. Moreover, it offers the advantage of covariantizing conventional coordinate calculations in GR. In the framework of symmetric teleparallel gravity, the associated energy-momentum density is fundamentally the Einstein pseudotensor, transformed into a true tensor. In the context of gravitational actions containing non-metricity, the action $S_{STEGR}=(-1/2k)\int\sqrt{-g}Q d^4x + S_m$, which substitutes the curvature scalar with the non-metricity, is at the basic of a theory called the Symmetric Teleparallel Equivalent of General Relativity (STEGR) \cite{heisenberg2024review}. 
The extension of symmetric teleparallel gravity led to the formulation of the $f(Q)$ gravity theory, also known as coincidence general relativity \cite{jimenez2018coincident} or non-metric gravity. In this theory, the connection is flat and torsionless. These conditions lead to a connection that is purely inertial, differing from the Levi-Civita connection through a general linear gauge transformation. Furthermore, the torsionless condition simplifies the connection to $Y^\alpha_{\;\;\mu\beta}=(\partial x^\alpha/\partial\xi^\lambda)\partial_\mu\partial_\beta\xi^\lambda$ for some arbitrary $\xi^\lambda$. This crucial outcome indicates that the connection can be entirely removed through a diffeomorphism. Consequently, the $\xi^\lambda$ fields emerge as St\"{u}ckelberg fields, restoring this gauge symmetry \cite{jimenez2018coincident}.
\par In exploring extensions of symmetric teleparallel gravity, recent studies have considered the characteristics of gravitational wave propagation. An analysis of the speed and polarization of gravitational waves \cite{soudi2019polarization} has remarkably extended the results obtained in general relativity, unveiling consistent speeds and polarizations.
\par In another line of research, in \cite{conroy2018spectrum}, a derivation of the exact propagator for the most general infinite-derivative, even-parity and generally covariant theory within symmetric teleparallel spacetimes was presented. This approach involves decomposing the action, containing the non-metricity tensor and its contractions, into terms involving the metric and a gauge vector field.
\par Further insights emerged from the study of new general relativistic type solutions in symmetric teleparallel gravity theories \cite{hohmann2019propagation}. The investigation of the gravitational wave propagation in Minkowski spacetime revealed that all gravitational waves propagate at the speed of light. The Noether symmetry approach played a key role in classifying possible first-order quadratic derivative terms of the non-metricity tensor in the framework of symmetric teleparallel geometry \cite{dialektopoulos2019noether}. The cosmology of the $f(Q)$ theory and its observational constraints were considered in \cite{lu2019cosmology} and \cite{lazkoz2019observational}, where it was shown that the accelerating expansion is an intrinsic property of the Universe's geometry, thus eliminating the need for exotic dark energy or additional fields and also used a dynamical system approach. For more work, check the Refs. \cite{d2022black,atayde2021can,Sokoliuk:2023ccw,Arora:2022mlo}.

\par Investigation of cosmological perturbations in $f(Q)$ gravity \cite{jimenez2020cosmology} revealed intriguing findings, such as the re-scaling of the Newton constant in tensor perturbations and the absence of vector contributions without vector sources being present. Notably, the scalar sector introduced two additional propagating modes, suggesting that $f(Q)$ theories add at least two extra degrees of freedom. Moreover, extending non-metric gravity by incorporating the trace of the matter-energy-momentum tensor $T$ into a general function $f(Q,T)$ has been investigated in \cite{xu2019f,xu2020weyl}. These $f(Q,T)$ gravity models have been observationally constrained as noted in \cite{arora2020f}, and some models have successfully described the accelerated expansion of the Universe \cite{arora2021constraining}. Additionally, a spherically symmetric stellar system in $f(Q,T)$ gravity has been shown to satisfy all the physical conditions \cite{das2024spherically}. For more works in $f(Q,T)$ gravity, see Refs. \cite{najera2022cosmological,najera2021fitting,yang2021geodesic,bhattacharjee2020baryogenesis,gadbail2021power,Arora:2021jik,Gadbail:2023klq}. Over the past two decades, numerous studies have been devoted to the geometrical and physical aspects of symmetric teleparallel gravity, with a surge in interest in recent years \cite{adak2006lagrange, adak2006symmetric, pala2022novel, adak2013symmetric, jimenez2016spacetimes, golovnev2017covariance, mol2017non, adak2018gauge, soudi2019polarization, conroy2018spectrum, jarv2018nonmetricity, delhom2018observable, harko2018coupling, hohmann2019propagation, dialektopoulos2019noether, lu2019cosmology, lazkoz2019observational, lobo2019novel, jimenez2020cosmology, beltran2019geometrical}.
\par Riemannian geometry represents a specific case within the broader framework of metric-affine geometry, offering a restricted perspective on gravitational dynamics. However, there exist no definitive physical principles that exclusively favor Riemannian geometry as the sole representation of gravity. Instead, metric-affine geometry presents three distinct yet physically equivalent avenues for describing gravitational phenomena. These approaches attribute the gravitational effects to the presence of non-zero curvature, non-zero torsion, or non-zero non-metricity within a given geometric framework. Together, these descriptions constitute the geometric trinity of GR \cite{heisenberg2019systematic, beltran2019geometrical}. It is essential to investigate all three approaches equally to gain a comprehensive understanding of gravity.
\par The coupling between the gravitational field and matter fields defines the dynamics in spacetime. In GR, the minimal coupling principle dictates that matter theories formulated in flat Minkowski space are seamlessly extended to incorporate gravitational interactions by replacing the flat metric and partial derivatives with the curved metric and covariant derivatives. This principle holds as long as the matter fields are coupled solely to the metric and its determinant without involving derivatives of the metric. In teleparallel gravity, for the electromagnetic potential, the presence of torsion introduces additional terms in the Maxwell action, violating the expected equivalence with GR \cite{jimenez2020coupling}. Similarly, fermionic fields are affected by torsion, further challenging the minimal coupling principle. However, in symmetric teleparallel gravity, the scenario shifts. The minimal coupling principle remains intact even in the presence of non-metricity \cite{jimenez2020coupling}. For electromagnetic fields and fermions alike, non-metricity does not interfere with the standard coupling prescriptions, ensuring compatibility with GR. In essence, while the symmetric teleparallel theory maintains equivalence with GR in the presence of matter fields, teleparallel theories diverge from this equivalence, underscoring the nuanced interplay between gravity and matter within these distinct frameworks.
\par The flat $\Lambda$CDM model generally aligns well with observations, but recent data indicate possible discrepancies. These include variations in the measured values of the Hubble constant $H_0$, and the amplitude of matter fluctuations $\sigma_8$, when different methods are used. Additionally, some anomalies arise when comparing the model's theoretical predictions, based on the best-fit cosmological parameters, with actual observations. These potential inconsistencies encourage the investigation of extensions of the  $\Lambda$CDM model. The well-known discrepancy between $H_0$, measured by the SH0ES collaboration using local distance ladder measurements from Type Ia supernovae $(H_0 = 73 \pm 1\,{\rm km/s/Mpc})$ \cite{riess2022comprehensive,murakami2023leveraging}, and the value inferred by the Planck collaboration from observations of temperature and polarization anisotropies in the Cosmic Microwave Background (CMB) radiation, assuming a $\Lambda$CDM cosmology $(H_0 = 67.4 \pm 0.5\,{\rm km/s/Mpc})$ \cite{aghanim2020planck}, has reached a statistical significance exceeding $5\sigma$. Unless this discrepancy is due to systematic errors, an intriguing possibility is that the Hubble tension could indicate new physics beyond the standard $\Lambda$CDM model of cosmology.
\par The primary aim of this study is to extend symmetric teleparallel gravity by incorporating the matter Lagrangian into the Lagrangian density of the $f(Q)$ theory \cite{jimenez2020cosmology}, thereby formulating the more general $f(Q,\mathcal{L}_m)$ framework. This approach allows for both minimal and non-minimal couplings between geometry and matter. Starting from the fundamental action of the model, we derive the general field equations by varying the action with respect to the metric. We also address the conservation properties of the matter energy–momentum tensor and demonstrate that, within this theory, it is generally non-conserved due to the geometry–matter coupling. Furthermore, for scenarios involving scalar fields, we obtain the corresponding modified Klein–Gordon equation, capturing the effects of the extended coupling on scalar field dynamics. The cosmological implications are then explored in the context of a flat Friedmann–Lemaître–Robertson–Walker (FLRW) metric, beginning with the derivation of the generalized Friedmann equations. Specific realizations of the theory, corresponding to distinct functional forms of $f(Q,\mathcal{L}_m)$, are examined in detail. The predictions of these models are confronted with independent observational datasets, enabling a direct comparison with standard cosmology.  Our findings reveal the rich and complex dynamics that can emerge in the Universe within this extended gravitational framework.

\par The paper is organized as follows. Section~\ref{sec: field_eq} presents the field equations of the generalized $f(Q,\mathcal{L}_m)$ gravity, beginning with the geometric preliminaries (subsection \ref{sec: geometric_basics}) and the variational principle in the metric formalism (subsection \ref{sec: variation_principle}). The corresponding Klein–Gordon equation is derived in subsection \ref{sec: cov}, and the energy–momentum tensor balance equations are discussed in subsection \ref{sec: momentum}. Section~\ref{sec: Cosmic_evolution} investigates the cosmological evolution of a flat FLRW universe in this framework, including the modified Friedmann equations and the de Sitter solution. The datasets and methodology of our MCMC analyses are summarized in Section~\ref{data}. Specific cosmological models, corresponding to different functional forms of $f(Q,\mathcal{L}_m)$, are analyzed in Section~\ref{sec: models}, where statistical results, correlations, and the effective mass of scalar field particles are also discussed. Finally, Section~\ref{sec: discussion} summarizes our main results and implications, while Appendix \ref{sec:Friedmann equations} contains the detailed derivation of the modified Friedmann equations.

\section{\texorpdfstring{Field equations of $f(Q,\mathcal{L}_m)$ gravity}{}} \label{sec: field_eq}

This section first provides a concise overview of the geometric foundations underlying gravitational theories, which are based on the existence of a general line element in spacetime. We then introduce the action of the $f(Q,\mathcal{L}_m)$ gravitational theory and, using the variational principle, derive the corresponding gravitational field equations, offering new insights into the role of geometry–matter couplings. The Klein–Gordon equation for scalar fields in this framework is also obtained, extending the analysis to field dynamics beyond the metric sector. Finally, we examine the non-conservation of the matter energy–momentum tensor, highlighting the direct impact of the coupling between the matter Lagrangian and geometry.

\subsection{Geometric Preliminaries}\label{sec: geometric_basics}

Once the definition of a metric is provided, the geometric interpretation of gravity is given by the Riemann tensor
\begin{equation}
    R^\alpha_{\;\;\beta\mu\nu}=\partial_{\mu}Y^\alpha_{\;\;\nu\beta} - \partial_{\nu}Y^\alpha_{\;\;\mu\beta}+ Y^\alpha_{\;\;\mu\lambda}Y^\lambda_{\;\;\nu\beta} - Y^\alpha_{\;\;\nu\lambda}Y^\lambda_{\;\;\mu\beta},
\end{equation}
and of its contractions. The Riemann tensor is constructed with the help of an affine connection. The general form of the affine connection $Y^\alpha_{\;\;\mu\nu}$ consists of three parts: a symmetric part known as the Levi-Civita connection $\Gamma^\alpha_{\;\;\mu\nu}$, a contortion tensor $K^\alpha_{\;\;\mu\nu}$ describing the anti-symmetric part, and the disformation tensor $L^\alpha_{\;\;\mu\nu}$, accounting for the presence of non-metricity,
\begin{equation}
Y^\alpha_{\;\;\mu\nu}=\Gamma^\alpha_{\;\;\mu\nu}+K^\alpha_{\;\;\mu\nu}+L^\alpha_{\;\;\mu\nu}.
\end{equation}
The torsion-free Levi-Civita connection $\Gamma^\alpha_{\;\;\mu\nu}$ is equivalent to the 2nd order Christoffel symbol in terms of the metric; it preserves the inner product of the various tangent vectors when a vector is parallelly transported, and it is defined according to
\begin{equation}
    \Gamma^\alpha_{\;\;\mu\nu}=\frac12 g^{\alpha\lambda}(\partial_\mu g_{\lambda \nu}+\partial_\nu g_{\lambda \mu} - \partial_\lambda g_{\mu\nu}).
\end{equation}
The contortion tensor $K^\alpha_{\;\;\mu\nu}$ is represented in terms of torsion tensor $T^\alpha_{\;\;\mu\nu}$ as
\begin{equation}
    K^\alpha_{\;\;\mu\nu}=\frac{1}{2}(T^\alpha_{\;\;\mu\nu}+ T_{\mu\;\;\nu}^{\;\;\alpha}+T_{\nu\;\;\mu}^{\;\;\alpha}).
\end{equation}
The torsion tensor characterizes the deviation of a connection from symmetry, which indicates that parallel transport around a closed loop does not necessarily bring a vector back to its original position.

The disformation tensor $L^\alpha_{\;\;\mu\nu}$ describes the general expansion or contraction of spacetime. When a vector is parallelly transported, its magnitude changes along its path. The variation of the length is measured by the non-metricity tensor,
\begin{equation}
    L^\alpha_{\;\;\mu\nu}=\frac{1}{2}(Q^\alpha_{\;\;\mu\nu}-Q^{\;\;\alpha}_{\mu\;\;\nu}-Q^{\;\;\alpha}_{\nu\;\;\mu}).
\end{equation}
The non-metricity tensor $ Q_{\alpha\mu\nu}$ is defined according to
\begin{equation}
    Q_{\alpha\mu\nu}=\nabla_\alpha g_{\mu\nu}= \partial_\alpha g_{\mu\nu} - Y^\beta_{\;\;\alpha\mu}g_{\beta\nu}-Y^\beta_{\;\;\alpha\nu}g_{\mu\beta}.
\end{equation}
To construct a boundary term in the action of the metric-affine gravity theories, we need a non-metricity conjugate, known as the superpotential $P^\alpha_{\;\;\mu\nu}$, defined as \cite{xu2019f}
\begin{equation}
    P^\alpha_{\;\;\mu\nu}= -\frac{1}{2}L^\alpha_{\;\;\mu\nu}+\frac{1}{4}(Q^\alpha-\Tilde{Q}^\alpha)g_{\mu\nu}-\frac{1}{4}\delta^\alpha_{\;\;(\mu}Q_{\nu)}.
\end{equation}
Here, $Q^\alpha=Q^{\alpha\;\;\mu}_{\;\;\mu}$ and $\Tilde{Q}^\alpha=Q_{\mu}^{\;\;\alpha\mu}$ are the non-metricity vectors. The non-metricity scalar can be obtained by contracting the superpotential tensor with the non-metricity tensor,
\begin{equation}
    Q=-Q_{\lambda\mu\nu}P^{\lambda\mu\nu}.
    \label{eq:non-metricityscalar}
\end{equation}
The non-metricity scalar $Q$ describes the deviation of the manifold geometry from isotropy and can be thought of as a measure of how much the volume of a parallelly transported object changes as it moves through spacetime.

\subsection{The variational principle and the field equation}\label{sec: variation_principle}

The dynamics of a physical system is studied using the action principle. The action for the  $f(Q,\mathcal{L}_m)$ modified gravity takes the following form
\begin{equation}
    S=\int f(Q,\mathcal{L}_m) \sqrt{-g} d^4x, \label{eq:Action}
\end{equation}
where $\sqrt{-g}$ is the determinant of the metric, and $f(Q,\mathcal{L}_m)$ is an arbitrary function of non-metricity scalar $Q$ and of the matter Lagrangian $\mathcal{L}_m$.\\
By varying the action with respect to the metric tensor, we obtain the gravitational field equation, which describes how spacetime geometry responds to the presence of matter and energy. Hence, we first obtain
\begin{equation}
    \delta S = \int \bigr[(f_Q \delta Q + f_{\mathcal{L}_m}\delta \mathcal{L}_m)\sqrt{-g}+f\delta\sqrt{-g} \bigl]d^4x.
    \label{eq:deltaofaction}
\end{equation}
Here, $f_Q=\partial f(Q,\mathcal{L}_m)/\partial Q$ and $f_{\mathcal{L}_m}=\partial f(Q,\mathcal{L}_m)/\partial \mathcal{L}_m$.\\
The variation of $Q$ is given by \cite{xu2019f}
\begin{equation}
    \delta Q= 2P_{\alpha\nu\rho}\nabla^\alpha\delta g^{\nu\rho}-(P_{\mu\alpha\beta}Q_\nu^{\;\;\alpha\beta}-2Q^{\alpha\beta}_{\;\;\;\mu}P_{\alpha\beta\nu})\delta g^{\mu\nu}\label{eq:delta_Q}.
\end{equation}
The energy-momentum tensor $T_{\mu\nu}$ of the matter is defined as \cite{landau2013classical}
\begin{equation}
    T_{\mu\nu}=-\frac{2}{\sqrt{-g}}\frac{\delta(\sqrt{-g}\mathcal{L}_m)}{\delta g^{\mu\nu}}=g_{\mu\nu}\mathcal{L}_m-2\frac{\partial \mathcal{L}_m}{\partial g^{\mu\nu}}.
    \label{eq:energy_tensor}
\end{equation}
The variation of the determinant of the metric tensor is
\begin{equation}
    \delta\sqrt{-g}=-\frac{1}{2}\sqrt{-g}g_{\mu\nu}\delta g^{\mu\nu}.
    \label{eq:delta_determinant}
\end{equation}
From Eqs. \eqref{eq:delta_Q}, \eqref{eq:energy_tensor}, and \eqref{eq:delta_determinant} it follows that Eq. \eqref{eq:deltaofaction} can be written as
\begin{equation}
    \begin{aligned}
    \delta S = \int \bigr[\bigr(f_Q ( 2P_{\alpha\nu\rho}\nabla^\alpha\delta g^{\nu\rho}-(P_{\mu\alpha\beta}Q_\nu^{\;\;\alpha\beta}-2Q^{\alpha\beta}_{\;\;\;\mu}P_{\alpha\beta\nu})\delta g^{\mu\nu}) &\\+ \frac{1}{2}f_{\mathcal{L}_m}(g_{\mu\nu}\mathcal{L}_m-T_{\mu\nu})\delta g^{\mu\nu}\bigl)-\frac{1}{2} f g_{\mu\nu}\delta g^{\mu\nu}\bigl]\sqrt{-g}d^4x.\label{eq:delta_penultimate_action}
    \end{aligned}
\end{equation}
After applying the boundary conditions and integrating the first term in Eq. \eqref{eq:delta_penultimate_action} becomes $-2\nabla^\alpha(f_Q\sqrt{-g}P_{\alpha\mu\nu})\delta g^{\mu\nu}$. Equating the metric variation of the action to zero,  we obtain the field equation of $f(Q,\mathcal{L}_m)$ gravity,
\begin{multline}
        \frac{2}{\sqrt{-g}}\nabla_\alpha(f_Q\sqrt{-g}P^\alpha_{\;\;\mu\nu}) +f_Q(P_{\mu\alpha\beta}Q_\nu^{\;\;\alpha\beta}-2Q^{\alpha\beta}_{\;\;\;\mu}P_{\alpha\beta\nu})
        + \\ \frac{1}{2}f g_{\mu\nu} = 
        \frac{1}{2}f_{\mathcal{L}_m}(g_{\mu\nu}\mathcal{L}_m-T_{\mu\nu}).\label{eq:fieldequation}
\end{multline}
For $f(Q,\mathcal{L}_m) = f(Q) + 2\,\mathcal{L}_m$, it reduces to the field equation of $f(Q)$ gravity (as seen in \cite{wang2022static})
\begin{equation}
    \frac{2}{\sqrt{-g}}\nabla_\alpha(\sqrt{-g}f_Q\,P^\alpha_{\;\;\mu\nu}) + f_Q\,q_{\mu\nu}+\frac{1}{2}f(Q) g_{\mu\nu}=-T_{\mu\nu},
\end{equation}
where $q_{\mu\nu}=P_{\mu\alpha\beta}Q_\nu^{\;\;\alpha\beta}-2Q^{\alpha\beta}_{\;\;\;\mu}P_{\alpha\beta\nu}$. Furthermore, the field equation \eqref{eq:fieldequation} can also be reduced to the STEGR.\\
In the mixed tensor representation, the field equation \eqref{eq:fieldequation} is given by
    \begin{multline}
    \frac{2}{\sqrt{-g}}\nabla_\alpha(f_Q\sqrt{-g}P^{\alpha\mu}_{\;\;\;\;\nu}) +f_Q\,P^{\mu}_{\;\;\alpha\beta}Q_\nu^{\;\;\alpha\beta}+\frac{1}{2}\delta^\mu_{\;\;\nu}\,f \\ =\frac{1}{2}f_{\mathcal{L}_m}(\delta^\mu_{\;\;\nu}\mathcal{L}_m-T^\mu_{\;\;\nu}).\label{eq:mixed_field_equation}
    \end{multline}
Using the Lagrange multiplier method with constraints $T^{\alpha}_{\;\;\beta\gamma}=0$ and $R^{\alpha}_{\;\;\beta\mu\nu}=0$, the action \eqref{eq:Action} reads as
\begin{equation}
    \begin{aligned}
         S=\int \bigr[ f(Q,\mathcal{L}_m) \sqrt{-g}  + \lambda_{\alpha}^{\;\;\beta\gamma}\,T^{\alpha}_{\;\;\beta\gamma} + \xi_{\alpha}^{\;\;\beta\mu\nu}\,R^{\alpha}_{\;\;\beta\mu\nu} \bigl] d^4x. \label{eq:lagrange_action}
    \end{aligned}
\end{equation}
The variation of the Lagrange multipliers is given as
\begin{equation}
    \delta( \lambda_{\alpha}^{\;\;\beta\gamma}\,T^{\alpha}_{\;\;\beta\gamma})=2\, \lambda_{\alpha}^{\;\;\beta\gamma}\,\delta Y^\alpha_{\;\;\beta\gamma},
\end{equation}
\begin{align}
    \delta( \xi_{\alpha}^{\;\;\beta\mu\nu} \,R^{\alpha}_{\;\;\beta\mu\nu}) & = \xi_{\alpha}^{\;\;\beta\mu\nu}\big[ \nabla_\mu(\delta\,Y^\alpha_{\;\;\nu\beta})-\nabla_\nu(\delta\,Y^\alpha_{\;\;\mu\beta})\big]
    \\ & = 2\,\xi_{\alpha}^{\;\;\nu\beta\mu}\nabla_\beta(\delta Y^\alpha_{\;\;\mu\nu}) \simeq 2(\nabla_\beta \xi_{\alpha}^{\;\;\nu\beta\mu})\delta Y^\alpha_{\;\;\mu\nu}.
\end{align}
Varying now the action \eqref{eq:lagrange_action} with respect to the connection gives
\begin{equation}
    \delta\,S = \int \left( 4\sqrt{-g}\,f_Q\,P^{\mu\nu}_{\;\;\;\;\alpha}+H_\alpha^{\;\;\mu\nu} + 2\,\nabla_\beta\xi_{\alpha}^{\;\;\nu\beta\mu}  
    +  2\, \lambda_{\alpha}^{\;\;\mu\nu}\right)d^4x\,\delta Y^\alpha_{\;\;\mu\nu}.
\end{equation}
Here $H_\alpha^{\;\;\mu\nu}$ is the hypermomentum density defined as
\begin{equation}
    H_\alpha^{\;\;\mu\nu}=\sqrt{-g}f_{\mathcal{L}_m}\frac{\delta \mathcal{L}_m}{\delta Y^\alpha_{\;\;\mu\nu}}.
\end{equation}
In the action variation, we introduce two covariant derivatives $\nabla_\mu\nabla_\nu$ to
eliminate the Lagrange multiplier coefficients with the anti-symmetry property of $\mu$ and $\nu$. Then, the field equation becomes
\begin{equation}
    \nabla_\mu\nabla_\nu\Bigl( 4\sqrt{-g}\,f_Q\,P^{\mu\nu}_{\;\;\;\;\alpha}+H_\alpha^{\;\;\mu\nu}\Bigl)=0.\label{eq:FE_connection}
\end{equation}

\subsection{The Klein-Gordon equation}\label{sec: cov}

In order to investigate the microphysical implications of our model, we consider the case of a scalar field $\phi$, described by the Klein-Gordon equation. Moreover, we assume that the scalar field is not a cosmological one but acts only at the level of the processes describing the microscopic interactions. A particular (and simple example) in this respect is the Higgs field, which contributes to (or generates) the mass of the elementary particles \cite{Wang:2023suf}, and thus is hidden inside ordinary matter.  This means that the explicit, or direct, cosmological effects of the scalar field we are investigating are negligible, and all its macroscopic effects can be expressed via the ordinary matter Lagrangian density.

On the other hand, we cannot rule out a priori the effect of the gravitation on the microscopic scalar field, at the level of elementary particles, and mass generation, and the possibility of the non-minimal coupling between scalar field and gravity (geometry).  Hence, based on the above considerations, and since one goal of our approach is to investigate the effects of modified gravity at a microphysical level, in the following, we consider only the effects of modified gravity on the scalar field, and we neglect the effects of the microscopic fields on the cosmological evolution.

There are several methods to include a nonminimal geometry-matter coupling into the Klein-Gordon equation, one of the most fundamental equations in field theory. In its standard and simplest form, this equation is known to be invariant under the group of conformal transformations with and without a mass term \cite{Wald:1984rg}. However, the massive Klein-Gordon equation can be made conformally invariant by the simple addition of the term $\xi R\phi$, with $\xi=1/6$ \cite{Wald:1984rg}. A similar extra term also arises in quantum field theory when one introduces a counter-term in the Lagrangian, which renormalizes a theory with an interacting scalar field in curved spacetime \cite{Bunch:1980br}. 

Alternatively, the generalized Klein-Gordon equation in the presence of geometry-scalar field coupling can be obtained from the variational approach, in a Lagrangian framework, by assuming a non-minimal quadratic coupling of the scalar field to the Ricci scalar. The action of the scalar field nonminimally coupled to gravity can be written in a covariant form as  \cite{Rossi:2019lgt, Amendola:1990nn} 
\begin{equation}
S_\phi=-\frac{1}{2}\int{\left(\nabla _\alpha \phi \nabla ^\alpha \phi +\xi R\phi ^2+\frac{m_0^2}{2}\phi^2\right)\sqrt{-g}d^4x},
\end{equation}
where  $\phi$ is the scalar field, $\Box = \nabla _\mu \nabla ^\mu$, $R$ is the Ricci scalar, $m_0$ denotes the mass of the scalar field particle, and $\xi$ is a dimensionless coupling constant. 

Hence in the presence of a non-minimally coupled scalar field the Klein-Gordon equation is given by \cite{Rossi:2019lgt, Amendola:1990nn}
\begin{equation}
    \left(\Box +m_0^2+\xi R\right)\phi=0.\label{eq:KGshort}
\end{equation}
The non-minimal scalar field - gravity coupling significantly modifies the Klein-Gordon equation. These modifications may have important implications for the overall dynamics and evolution of the elementary particles under the influence of gravitational fields, and they may also lead to some observational/experimental consequences that may allow testing the presence and magnitude of the modified gravity effects.  

For $\xi=0$, Eq. \eqref{eq:KGshort} reduces to the standard form of the Klein-Gordon equation in the presence of a minimal coupling, $ \left(\Box +m_0^2\right)\phi=0$.

The $f(Q,\mathcal{L}_m)$-field Eq. \eqref{eq:fieldequation} can be rewritten in a covariant form, similar  to the standard Einstein gravitational field equations as (see \cite{Zhao:2021zab} for the detailed calculation),
\begin{equation}
\begin{aligned}
\label{eq:einsQLm}
    f_Q{G}_{\mu \nu}+\frac{1}{2} g_{\mu \nu}\left(f- f_Q Q\right) +  2 &f_{Q Q}\left(\partial_\alpha Q\right) P^\alpha{ }_{\mu \nu} \\
    & =\frac{1}{2}f_{\mathcal{L}_m}(g_{\mu\nu}\mathcal{L}_m-T_{\mu\nu}).
\end{aligned}
\end{equation}
In the STGR limit with $f(Q) = -Q + 2\mathcal{L}_m$, the left-hand side of Eq.~\eqref{eq:einsQLm} reduces to the Einstein tensor $G_{\mu \nu}$, which only depends on the metric of the spacetime manifold.

By introducing the notations,
\begin{equation}
    \Delta =\frac{f_{\mathcal{L}_m}}{f_{Q}}, \, \delta =\frac{f}{f_{\mathcal{L}_m}}, \, \Sigma=\frac{f}{f_Q},\Psi_\alpha=2\frac{f_{QQ}}{f_Q}\partial _\alpha Q,\label{eq:notation}
\end{equation}
Eq.~\eqref{eq:einsQLm} can be reformulated as
\begin{equation}
G_{\mu \nu}+\frac{1}{2}g_{\mu \nu}\left(\Sigma -Q\right)+\Psi_\alpha P^\alpha _{\,\,\,\mu \nu}=\frac{1}{2}\Delta \left(g_{\mu \nu}\mathcal{L}_m-T_{\mu \nu}\right).
\end{equation}

Taking the trace of Eq. \eqref{eq:einsQLm}, and after systematic algebraic simplifications, we obtain the Ricci scalar of the $f(Q,\mathcal{L}_m)$ gravity as,
\begin{equation}
    R=\frac{1}{2}\Delta\left(T-4\mathcal{L}_m\right)+2\left(\Sigma-Q\right) + \partial_\alpha Q\left(Q^\alpha-\Tilde{Q^\alpha}\right)\frac{\partial}{\partial Q}\log{f_Q}.\label{eq:RinfQlm}
\end{equation}

Substituting Eq.~\eqref{eq:RinfQlm} into Eq.~\eqref{eq:KGshort} results in the modified Klein-Gordon equation in the $f(Q,\mathcal{L}_m)$ gravity,
\begin{equation}
    \left(\Box +m_\mathrm{eff}^2\right)\phi=0,
\end{equation}
where we have denoted,
\begin{equation}
\begin{aligned}
m_\mathrm{eff}^2 = m_0^2 
&+ \xi \Bigg[ \frac{1}{2} \Delta \left( T - 4\mathcal{L}_m \right) 
+ 2 \left( \Sigma - Q \right)  \\
&\qquad \qquad + \partial_\alpha Q \left( Q^\alpha - \tilde{Q}^\alpha \right) 
\frac{\partial}{\partial Q} \log f_Q \Bigg].
\label{eq:meff}
\end{aligned}
\end{equation}

Thus, $m_\mathrm{eff}$ represents the effective mass of the scalar field in $f(Q,\mathcal{L}_m)$ gravity. The scalar field interacts not only with its own mass $m_0$, but also with the non-metricity and matter lagrangian, as described by the additional terms in the equation. This kind of generalization of the Klein-Gordon equation allows for rich scalar field dynamics and can give some novel insights into the explanation of phenomena such as cosmic acceleration, inflation, or dark energy.

\subsection{Energy-momentum tensor balance equation}\label{sec: momentum}

The covariant derivative of $\omega^\mu_{\;\;\nu}$ is given by
\begin{equation}
    \nabla_\mu\,\omega^\mu_{\;\;\nu}=D_\mu\,\omega^\mu_{\;\;\nu}-\frac{1}{2}Q_\rho\,\omega^\rho_{\;\;\nu} - L^{\lambda}_{\;\;\mu\nu}\,\omega^\mu_{\;\;\lambda},\label{eq: covariant_derivative}
\end{equation}
where $D_\mu$ is the covariant derivative with respect to the Levi-Civita connection. The covariant derivative of the field equation \eqref{eq:mixed_field_equation} is
\begin{equation}
\begin{aligned}
        D_\mu\Bigr[ \frac 12 f_{\mathcal{L}_m}(\delta^\mu_\nu\,\mathcal{L}_m -  T^\mu_{\;\;\nu})\Bigl]= \frac{1}{2}&\partial_\nu f+ D_\mu (f_Q P^\mu_{\;\;\alpha\beta}Q_\nu^{\;\;\alpha\beta}) \\ & + D_\mu\Bigr[ \frac{2}{\sqrt{-g}}\nabla_\alpha(f_Q\sqrt{-g}P^{\alpha\mu}_{\;\;\;\;\nu}) \Bigr].\label{eq:Cov_derof_FE}
\end{aligned}
\end{equation}
From Eq. \eqref{eq: covariant_derivative} $D_\mu$ can be expressed as $D_\mu=\nabla_\mu+\frac{1}{2}Q_\mu + L^{\rho}_{\;\;\mu\nu}$. Thus, Eq. \eqref{eq:Cov_derof_FE} becomes
\begin{widetext}

\begin{equation}
\begin{aligned}
    \frac{1}{2\sqrt{-g}} \nabla_\alpha\nabla_\mu H_\nu^{\;\;\alpha\mu} - \frac{1}{2}f_{\mathcal{L}_m}D_\mu\,T^\mu_{\;\;\nu}= &\frac{1}{2}f_Q\,\partial_\nu Q \ +  \nabla_\mu(f_Q P^\mu_{\;\;\alpha\beta}Q_\nu^{\;\;\alpha\beta}) \ + \frac{1}{2} Q_\mu (f_Q P^\mu_{\;\;\alpha\beta}Q_\nu^{\;\;\alpha\beta}) \ \\& + L^{\rho}_{\;\;\mu\nu}(f_Q P^\mu_{\;\;\alpha\beta}Q_\rho^{\;\;\alpha\beta}) + \frac{2}{\sqrt{-g}}L^{\rho}_{\;\;\mu\nu}\nabla_\alpha(\sqrt{-g}f_Q\,P^{\alpha\mu}_{\;\;\;\;\rho}) 
    + \frac{1}{\sqrt{-g}}Q_\mu\nabla_\alpha(\sqrt{-g}f_Q\,P^{\alpha\mu}_{\;\;\;\;\nu}).
\end{aligned}
\end{equation}
\end{widetext}
The detailed calculations are shown in \cite{xu2019f}, and they lead to
\begin{eqnarray}\label{eq:Momentum_ordinary}
    D_\mu\,T^\mu_{\;\;\nu}= \frac{1}{f_{\mathcal{L}_m}\sqrt{-g}}\left[ \nabla_\alpha\nabla_\mu H_\nu^{\;\;\alpha\mu} - 2\,Q_\mu\,\nabla_\alpha(f_Q\sqrt{-g}P^{\alpha\mu}_{\;\;\;\;\nu}) \right].\nonumber
\end{eqnarray}
To simplify the above equation, we introduce the tensor $A^{\mu}_{\;\;\alpha}$ and define Eq. \eqref{eq:FE_connection} such that
\begin{equation}
     \nabla_\mu\Bigl( 4\sqrt{-g}\,f_Q\,P^{\mu\nu}_{\;\;\;\;\alpha}+H_\alpha^{\;\;\mu\nu}\Bigl)= \sqrt{-g}A^{\nu}_{\;\;\alpha}.\label{eq:A_assumed_tensor}
\end{equation}
Then the covariant derivative of the RHS of Eq.~(\ref{eq:A_assumed_tensor}) is
\begin{equation}\label{eq:cov_der_A}
  \nabla_\nu(\sqrt{-g}A^{\nu}_{\;\;\alpha})=\sqrt{-g}\nabla_\nu\,A^{\nu}_{\;\;\alpha}+\frac{\sqrt{-g}}{2}Q_\nu\,A^{\nu}_{\;\;\alpha}
     =0.
\end{equation}
 Eq. \eqref{eq:Momentum_ordinary} simplifies by the combination of Eqs. \eqref{eq:A_assumed_tensor} and \eqref{eq:cov_der_A} as
\begin{multline}\label{eq:momentum_conservation}
        D_\mu\,T^\mu_{\;\;\nu}= \frac{1}{f_{\mathcal{L}_m}}\left[ \frac{2}{\sqrt{-g}}\nabla_\alpha\nabla_\mu H_\nu^{\;\;\alpha\mu} + \nabla_\mu\,A^{\mu}_{\;\;\nu} - \right. \\  \left. \nabla_\mu \left( \frac{1}{\sqrt{-g}}\nabla_\alpha H_\nu^{\;\;\alpha\mu}\right) \right]=B_\nu \neq 0.
\end{multline}
From Eq.~\eqref{eq:momentum_conservation}, it follows that the matter energy-momentum tensor is not conserved in the $f(Q,\mathcal{L}_m)$ gravity theory. The non-conservation tensor $B_\nu$ is a function of dynamical variables like $Q$, $\mathcal{L}_m$, and the thermodynamic parameters.\par
In a broader context, dissipative processes pose significant challenges when reconciling cosmic microwave background radiation (CMBR) and large-scale structure (LSS). In \cite{shabani2014cosmological}, the cosmological and solar system consequences of a class of models with geometry-matter coupling were investigated. The findings of this work suggest that these models often exhibit inconsistent behavior compared to the observational data. This inconsistency may potentially manifest and amplify when extended to cosmological scales at both galactic and extragalactic levels. In particular, incompatibility with CMBR or LSS appears to be a model-dependent phenomenon. However, the study in \cite{shabani2014cosmological} reveals that some or all of these inconsistencies can be mitigated through meticulous fine-tuning of model parameters.\par
At larger scales, specifically galactic and extragalactic levels, the non-minimal matter coupling with geometry introduces intriguing implications. The observed flattening of galaxy rotation curves, considered a dynamically generated effect, is attributed to non-minimal coupling \cite{bertolami2010dark,paramos2010dark}. The non-conservation of the energy-momentum tensor leads to a deviation from geodesic motion, which explained the observed deviation between measured rotation velocity and classical predictions. Moreover, a specific type of non-minimal matter coupling with geometry is shown to mimic the presence of dark matter in galaxy clusters. In \cite{bertolami2012mimicking}, they explore this phenomenon in the context of the Abell cluster A586, demonstrating its potential extension to a larger sample of galaxy clusters. Adding to the complexity of the physical behavior, dissipative processes play a distinctive role in the evolution of radio galaxies, as discussed in \cite{perucho2019dissipative}.\par
If we consider matter as a perfect fluid described by its pressure $p$ and energy density $\rho$, the energy-momentum tensor can be defined as
\begin{equation}
    T^{\mu}_{\;\;\nu}=(\rho+p)u_\nu\,u^{\mu}+p\,\delta^{\mu}_{\nu},
\end{equation}
where $u^\mu$ denotes the four-velocity of the fluid. Following \cite{harko2018coupling} we have
\begin{equation}
    \dot{\rho} +3\,H(\rho+p)=B_\mu\,u^{\mu}.
\end{equation}

The continuity equation presented above deviates noticeably from the standard form, incorporating additional terms on the right-hand side (RHS) that account for the deviations from the geodesic motion. In this context, the source term, denoted by $B_\mu\,u^{\mu}$, is associated with the generation or dissipation of energy. When $B_\mu\,u^{\mu}=0$, the system obeys the energy conservation law of standard gravity. In contrast, if $B_\mu\,u^{\mu}$ takes nonzero values, energy transfer processes become dominant.\par

The momentum conservation equation, which describes the movement of massive particles \cite{xu2019f,harko2018coupling}, is expressed as
\begin{equation}
    \frac{d^2x^\mu}{ds^2}+\Gamma^\mu_{\alpha\beta}u^\alpha u^\beta = \frac{h^{\mu\nu}}{\rho+p}(B_\nu-D_\nu\,p)=F^{\mu},
\end{equation}
where $h^{\mu\nu}$ represents the projection operator, defined as $h^{\mu\nu}=g^{\mu\nu} + u^\mu u^\nu$. The equation of motion exhibits a notable departure from the geodesic motion of the massive particles. An additional force, $F^\mu$, emerges as a consequence of the coupling between $Q$ and $\mathcal{L}_m$. This coupling introduces a non-gravitational influence, leading to deviations from the trajectories determined by the standard geodesic motion of GR, thus influencing the dynamical evolution of massive particles.

\section{\texorpdfstring{Cosmological evolution of FLRW universe in $f(Q,\mathcal{L}_m)$ gravity}{}}\label{sec: Cosmic_evolution}

In the present Section, we will investigate, in a general framework,  the cosmological implications of the $f(Q,\mathcal{L}_m)$ gravity theory. By considering a flat Friedmann-Lemaitre-Robertson-Walker (FLRW) geometry, the generalized Friedmann equations are derived. The cosmological evolution equations do contain some extra terms, coming from the presence of the nonmetricity and geometry matter coupling, which generate an effective density of pressure, which can be interpreted as representing geometric dark energy. The general form of the energy balance equation is also obtained. The de-Sitter limiting behavior of the cosmological models is also investigated.

\subsection{The Friedmann equations}\label{sec: friedmann_derivation}
To study the cosmological evolution in $f(Q,\mathcal{L}_m)$ gravity, we assume that the Universe is described by FLRW geometry, with the spacetime interval of the form
\begin{equation}
    ds^2=-dt^2+a^2(t)(dx^2+dy^2+dz^2),
\end{equation}
where $a(t)$ is the scale factor\footnote{Here, we assume the Lapse function as $N(t)=1$.}. We define the rate of expansion of the universe as $H = \frac{\dot{a}}{a}$. Let us also assume that the Universe is filled with a perfect fluid. We adopt the expressions $\mathcal{L}_m = -\rho$, or $\mathcal{L}_m=p$ for the Lagrangian density of cosmic matter. Hence, in the comoving frame, the non-zero components of the energy-momentum tensor are given by $T^{\mu}_{\nu}=(-\rho,p,p,p)$.

Using the FLRW metric, the field equations Eq. \eqref{eq:fieldequation} give the two generalized Friedmann equations (the detailed calculations are presented in the Appendix \ref{sec:Friedmann equations})
\begin{eqnarray}
\label{eq:friedmann1}
    && 3H^2 =\frac{1}{4f_Q}\bigr[ f - f_{\mathcal{L}_m}(\rho + \mathcal{L}_m) \bigl],\\
     \label{eq:friedmannmain}
   && \dot{H} + 3H^2 + \frac{\dot{f_Q}}{f_Q}H=\frac{1}{4f_Q}\bigr[ f + f_{\mathcal{L}_m}(p - \mathcal{L}_m) \bigl]. 
\end{eqnarray}
For $f(Q,\mathcal{L}_m)=f(Q)+2\,\mathcal{L}_m$ the Friedmann equations reduces to $f(Q)$ \cite{solanki2021cosmic, mandal2020cosmography}, it can further be simplified to STEGR. By subtracting Eqs. \eqref{eq:friedmann1} and \eqref{eq:friedmannmain}, we obtain
\begin{equation}
\label{com}
    \frac{d}{dt}(f_Q\,H)=\frac{f_{\mathcal{L}_m}}{4}(p+\rho).
\end{equation}
The generalized expression of the deceleration parameter is obtained as
\begin{equation}
\begin{aligned}
    q =& -1 - \frac{\dot{H}}{H^{2}} \\
      =& \frac{1}{4 f_{Q}H^{2}} \left(2 Q f_{Q} + 4 \dot{f_{Q}} H - f - f_{\mathcal{L}_m}(p-\mathcal{L}_m)\right) - 1.
\end{aligned}
\end{equation}
With the help of Eq. \eqref{eq:friedmann1}, Eq. \eqref{eq:friedmannmain} can be rewritten as
\begin{equation}
2\dot{H}+3H^2=\frac{1}{4f_Q}\left[f+f_{\mathcal{L}_m}\left(\rho+2p-\mathcal{L}_m\right)\right]-2\frac{\dot{f}_Q}{f_Q}H.
\end{equation}
Thus, we can reformulate the generalized Friedmann equations of the $f\left(Q,\mathcal{L}_m\right)$ gravity theory in the form
\begin{equation}\label{41}
3H^2=\rho_{eff}, \;\;\;\;\;  2\dot{H}+3H^2=-p_{eff},
\end{equation}
where we have introduced the effective energy density and pressure, defined as
\begin{equation}
\rho_{eff}=\frac{1}{4f_Q}\bigr[ f - f_{\mathcal{L}_m}(\rho + \mathcal{L}_m) \bigl],
\end{equation}
and
\begin{equation}
p_{eff}=2\frac{\dot{f}_Q}{f_Q}H-\frac{1}{4f_Q}\left[f+f_{\mathcal{L}_m}\left(\rho+2p-\mathcal{L}_m\right)\right],
\end{equation}
respectively. Eq.~(\ref{41}) allow to formulate the generalized effective conservation equation of the $f\left(Q,\mathcal{L}_m\right)$ gravity theory as
\begin{equation}\label{cons1}
\dot{\rho}_{eff}+3H\left(\rho_{eff}+p_{eff}\right)=0.
\end{equation}
Using Eq. \ref{eq:notation} we can represent the effective energy density and pressure as
\begin{equation}
\rho _{eff}=\frac{1}{4}\Delta \left[ \delta -\left( \rho +\mathcal{L}_m\right) %
\right] ,
\end{equation}%
and
\begin{equation}
p_{eff}=2\frac{\dot{f}_{Q}}{f_{Q}}H-\frac{1}{4}\Delta \left[ \delta +\left(
\rho +2p-\mathcal{L}_m\right) \right] ,
\end{equation}%
respectively.  Then the conservation equation (\ref{cons1}) can be
reformulated as
\begin{multline}
\dot{\rho} + 3H\left( \rho +p\right) =\frac{1}{\Delta }\frac{d}{dt}\left[
\Delta \left( \delta -\mathcal{L}_m\right) \right]+ \\ 3H\left\{ 8\frac{\dot{f}_{Q}}{
f_{Q}}\frac{H}{\Delta }-\left[ \left( 1+\frac{\dot{\Delta}}{\Delta }\right)
\rho +p\right] \right\} =\Gamma.
\end{multline}
The function $\Gamma $ describes the non-conservation level of the present
modified gravity theory. If $\Gamma >0$, the energy of the particles
increases due to the energy transfer of matter to the gravitational field.
The case $\Gamma <0$ can be interpreted as describing particle decay due to
the matter-geometry coupling.

From Eqs.~(\ref{41}) we also obtain the expression of the deceleration parameter as
\begin{equation}
\begin{aligned}
q=&\frac{1}{2}+\frac{3}{2}\frac{p_{eff}}{\rho_{eff}}\\
=&\frac{1}{2}+6\frac{2\dot{f}_QH-(1/4)\left[f+f_{\mathcal{L}_m}\left(\rho+2p-\mathcal{L}_m\right)\right]}{f-f_{\mathcal{L}_m}\left(\rho+\mathcal{L}_m\right)}.
\end{aligned}
\end{equation}

The Universe enters an accelerating phase when $q<0$, or $p_{eff}<-\rho_{eff}/3$. This gives the condition that must be satisfied by the function $f$ and its derivatives to describe an accelerated expansion
\begin{equation}
12\dot{H}f_Q+f-f_{\mathcal{L}_m}\left(\rho+\mathcal{L}_m\right)>0.
\end{equation}

To compare the theoretical results with the cosmological observations, we introduce an independent variable redshift $z$ instead of the usual time
variable $t$, defined as $a = 1/(1+z)$, where we have used a normalization of the scale factor by imposing $a(0)=1$. Thus, we can replace the derivatives with respect to the time with the derivatives with respect to the redshift using the relation
\begin{eqnarray}
    \frac{d}{dt} = -(1+z) H(z) \frac{d}{dz}.
\end{eqnarray}
Moreover, the redshift dependence of the deceleration parameter is given by
\begin{eqnarray}
    q(z) = -1 + (1+z) \frac{H'(z)}{H(z)}.
\end{eqnarray}

\subsection{The de Sitter solution}\label{sec: De-sitter_soln}

As a first step in considering explicit theoretical models, we consider the problem of the existence of a de-Sitter type vacuum solution of the cosmological field equations. The de Sitter solution corresponds to $p=0$, $\rho=0$ and $H=H_{0}={\rm constant}$, respectively. For a vacuum de-Sitter type Universe, Eq.~\eqref{com} gives $\dot{f_{Q}}=0$, and further results in $f_{Q}=F_{0}$, where $F_{0}$ is a constant.

The condition $f_{Q}=F_{0}$ is satisfied for any $Q$, when we have \cite{xu2019f,harko2018coupling}
\begin{eqnarray}
    f(Q)= F_{0}Q+ 2\Lambda,
\end{eqnarray}
where $\Lambda$ is an integration constant. In the vacuum de Sitter phase, the first field equation \eqref{eq:friedmann1} reduces to the form
\begin{eqnarray}
    3 H_{0}^2 = \frac{6F_{0}H_{0}^{2}+2\Lambda}{4 F_{0}}.
\end{eqnarray}
One can also write the above equation as
\begin{eqnarray}
    H_{0}=\sqrt{\frac{\Lambda}{3F_{0}}}.
\end{eqnarray}
Hence, the $f(Q,\mathcal{L}_m)$ theory admits the de-Sitter type evolution in the limiting case of a vacuum Universe. As can be easily calculated, for the de-Sitter solution, we have $q =-1$ and $\omega =-1$, respectively.

\section{Data and methodology of MCMC analyses} 
\label{data}

This section outlines the observational datasets employed to constrain the $f(Q, \mathcal{L}_m)$ modified gravity model using a Markov Chain Monte Carlo (MCMC) approach. A Bayesian statistical analysis is carried out with the \texttt{emcee} MCMC sampler \cite{Foreman-Mackey:2012any}, implemented within the Python environment, to derive credible bounds on the model parameters. To maximize the likelihood function for each dataset, we adopt the following flat priors: $H_0 \in [30, 100]$, $\alpha \in [0, 0.5]$, $\beta \in [-1.1, -0.2] \times 10^{4}$, $n \in [-2,2]$, and $\gamma \in [0, 2]$. Additionally, a joint analysis is performed by combining the observational samples from cosmic chronometers, Type Ia supernovae, and baryon acoustic oscillations.

To gain deeper insights into the nature of dark energy, the analysis incorporates the following observational datasets:
\begin{itemize}
\item Cosmic Chronometers (CC): This dataset consists of Hubble parameter measurements obtained using the cosmic chronometer approach \cite{Yu_2018}. The Hubble function $H(z)$ is determined by evaluating the derivative of cosmic time with respect to redshift, i.e., $H(z) = -\frac{1}{1+z} \frac{dz}{dt}$, at redshifts $z \neq 0$. This method relies on estimating the differential age, $\Delta t$, between passively evolving galaxies at slightly different redshifts to infer $\Delta z/\Delta t$, offering a model-independent probe of the Universe's expansion history.
		
\item Baryon Acoustic Oscillations (BAO): We employ BAO measurements from the second data release (DR2) of the Dark Energy Spectroscopic Instrument (DESI), which includes observations of galaxies, quasars, and Lyman-$\alpha$ forest tracers. These measurements span the redshift range $0.295 \leq z \leq 2.330$, divided into nine redshift bins, and provide both isotropic and anisotropic BAO constraints. The dataset reports measurements in terms of the transverse comoving distance $D_M/r_d$, the Hubble horizon $D_H/r_d$, and the angle-averaged distance $D_V/r_d$, all normalized by the comoving sound horizon at the drag epoch, $r_d$. This dataset is hereafter referred to as DESI-DR2 \citep{DESI:2025zgx,DESI:2024mwx}.
		
\item Type Ia Supernovae (SNeIa): We utilize the Pantheon+ (PP) compilation, which consists of 1701 light curves from 1550 distinct SNeIa events, covering the redshift range $0.01 \leq z \leq 2.26$ \citep{Brout_2022,riess2022comprehensive,Murakami_2023}. For this analysis, we exclude the SH0ES calibration and instead use the observed apparent magnitude values, denoted by $m$. The distance modulus is defined as $\mu \equiv m - M_b = 5 \log_{10}(D_L/\text{Mpc}) + 25$, where $M_b$ is the absolute magnitude of the SNeIa, and $D_{L}$ is the luminosity distance.

\end{itemize}

To perform the MCMC sampling, we minimize the combined chi-squared functions $\chi^{2}_{OHD} +\chi^{2}_{SN} +\chi^{2}_{BAO}$ which correspond to the log-likelihood defined as $\mathcal{L}=\exp(-\chi^2/2)$. The resulting best-fit parameter values, along with their 68\% confidence level uncertainties, are summarized in Table \ref{tab:model_params}.

To assess how the proposed models compare to the standard \(\Lambda\)CDM scenario, we use the Akaike Information Criterion (AIC) and Bayesian Information Criterion (BIC) \citep{1100705}, defined as
\begin{equation}
\mathrm{AIC} = \chi^2_{\mathrm{min}} + 2d, \quad \mathrm{BIC} = \chi^2_{\mathrm{min}} + d \ln N,
\end{equation}

where \(d\) is the number of free parameters and \(N\) is the total number of data points. In contrast, for the reduced chi-square statistic, the model most favored by the data is the one with a $\chi^2_\nu$ value closest to 1. A value much greater than $1$ indicates a poor fit, implying the model is underfitted, whereas a value much less than $1$ suggests overfitting, either because the model inadequately accounts for noise or because the error variance has been overestimated. To evaluate the relative performance of each model with respect to the standard \( \Lambda \)CDM cosmology, we compute the differences in the Akaike and Bayesian Information Criteria, defined as \( \Delta \mathrm{AIC} = \mathrm{AIC}_{\text{model}} - \mathrm{AIC}_{\Lambda \text{CDM}} \) and \( \Delta \mathrm{BIC} = \mathrm{BIC}_{\text{model}} - \mathrm{BIC}_{\Lambda \text{CDM}} \), respectively. A negative value of \( \Delta \mathrm{AIC} \) or \( \Delta \mathrm{BIC} \) indicates that the proposed model is statistically favored over \( \Lambda \)CDM, as it achieves a better trade-off between goodness of fit and model complexity, while a positive value implies that \( \Lambda \)CDM is preferred. 

 The interpretation of $\Delta X$ (X can be AIC or BIC) is as follows: if $0 \leq \Delta X \leq 2$ or $-2 \leq \Delta X < 0$, the evidence is weak and it is not possible to determine which model is preferred; if $2 < \Delta X \leq 6$ or $-6 \leq \Delta X < -2$, the evidence is positive; if $6 < \Delta X \leq 10$ or $-10 \leq \Delta X < -6$, the evidence is strong; and if $\Delta X > 10$ or $\Delta X < -10$, the evidence is very strong. A summary of the model comparisons is presented in Table~\ref{tab:stat}.

\section{Cosmological models}\label{sec: models}

In this Section, we will explore various cosmological models based on the $f(Q,\mathcal{L}_m)$ gravity theory. The models are determined by specific choices for the functional form of $f\left(Q,\mathcal{L}_m\right)$. To keep our analysis as general as possible, we will assume that the matter in the Universe obeys an equation of state given by $p=(\gamma-1)\rho$ where $p$ is the pressure and $\rho$ is the energy density, $0 \leq \gamma \leq 2$. For $\gamma =4/3$, this linear relationship between pressure and energy density describes the behavior of the radiation in the early Universe, characterized by high density, as well as, for $\gamma =1$, in the present Universe,  when the matter density is low.

The degeneracy of the matter Lagrangian in general relativity is a well-known issue, thoroughly discussed in the seminal work \cite{Brown:1992kc,deFelice:1990hu}, where this degeneracy was first analyzed in detail. These studies show that the matter Lagrangian is not an explicit function of the metric and may be chosen as either \( -\rho \) or \( p \), with both options yielding the same energy–momentum tensor in standard GR. However, in modified gravity theories with non-minimal coupling between geometry and matter, such as \( f(Q,\mathcal{L}_m) \) gravity, the explicit form of \( \mathcal{L}_m \) becomes physically relevant and must be specified to fully determine the dynamics. In this work, we adopt \( \mathcal{L}_m = p \), a choice frequently used in the literature, as it provides a consistent and mathematically tractable formulation, particularly suitable for cosmological applications within a FLRW background.

\subsection{Model I: \texorpdfstring{$f=-\alpha\, Q+ 2\mathcal{L}_m+\beta$ with $\mathcal{L}_m=p$}{}}\label{sec: model_A}

As a first example of a cosmological model, we consider the functional form
\[
f(Q, \mathcal{L}_m) = -\alpha\, Q + 2 \mathcal{L}_m + \beta\,,
\]
where \(\mathcal{L}_m = p\) denotes the matter Lagrangian equal to the pressure, and \(\alpha\) and \(\beta\) are constants. For this specific \(f(Q, \mathcal{L}_m)\) model, with \(\mathcal{L}_m = p = (\gamma - 1) \rho\), the Friedmann equations reduce to
\begin{eqnarray}
\label{bb1}
&& 3 H^{2}= -\frac{\beta}{2\alpha} + \frac{\rho}{\alpha},\\
\label{bb2}
 && 2 \,  \dot{H}+ 3  H^{2} = -3  H^{2}(\gamma-1)-\frac{\beta\gamma}{2\alpha}.
\end{eqnarray}
The system of Eqs.~(\ref{bb1}) and (\ref{bb2}) admits a de Sitter-type solution in a vacuum Universe, characterized by a constant Hubble parameter \(H = H_0\). This solution satisfies the relation $3 H_0^2 = -\frac{\beta}{2 \alpha}$, which implies that either \(\alpha\) or \(\beta\) must be negative. The corresponding effective energy densities and pressures are given by
\begin{equation}
\rho_{eff}=\frac{\rho}{\alpha}-\frac{\beta}{2\alpha}, \ p_{eff}= 3  H^{2}(\gamma-1) + \frac{\beta\gamma}{2\alpha},
\end{equation}
leading to the energy balance equation
\begin{equation}
\dot{\rho}+3H\left(\rho+p\right)=0,
\end{equation}
Hence, in this model, the energy-momentum tensor of matter is conserved.

By employing the relation \(\frac{1}{H} \frac{dH}{dt} = \frac{dH}{d \ln a}\), the equations admit an exact solution for the Hubble parameter, expressed as
\begin{equation}
\label{H1}
H(z) = \left[ \frac{(6 H_0^2 \alpha + \beta)(1+z)^{3\gamma} - \beta}{6 \alpha} \right]^{\frac{1}{2}},
\end{equation}
where \(H(0) = H_0\) denotes the present-day Hubble parameter. Note that the positivity condition on \(H(z)\) in Eq.~\eqref{H1} imposes constraints on the model parameters, which guide the choice of priors in the MCMC analysis.

\begin{figure}[]
\centering
\includegraphics[scale=0.42]{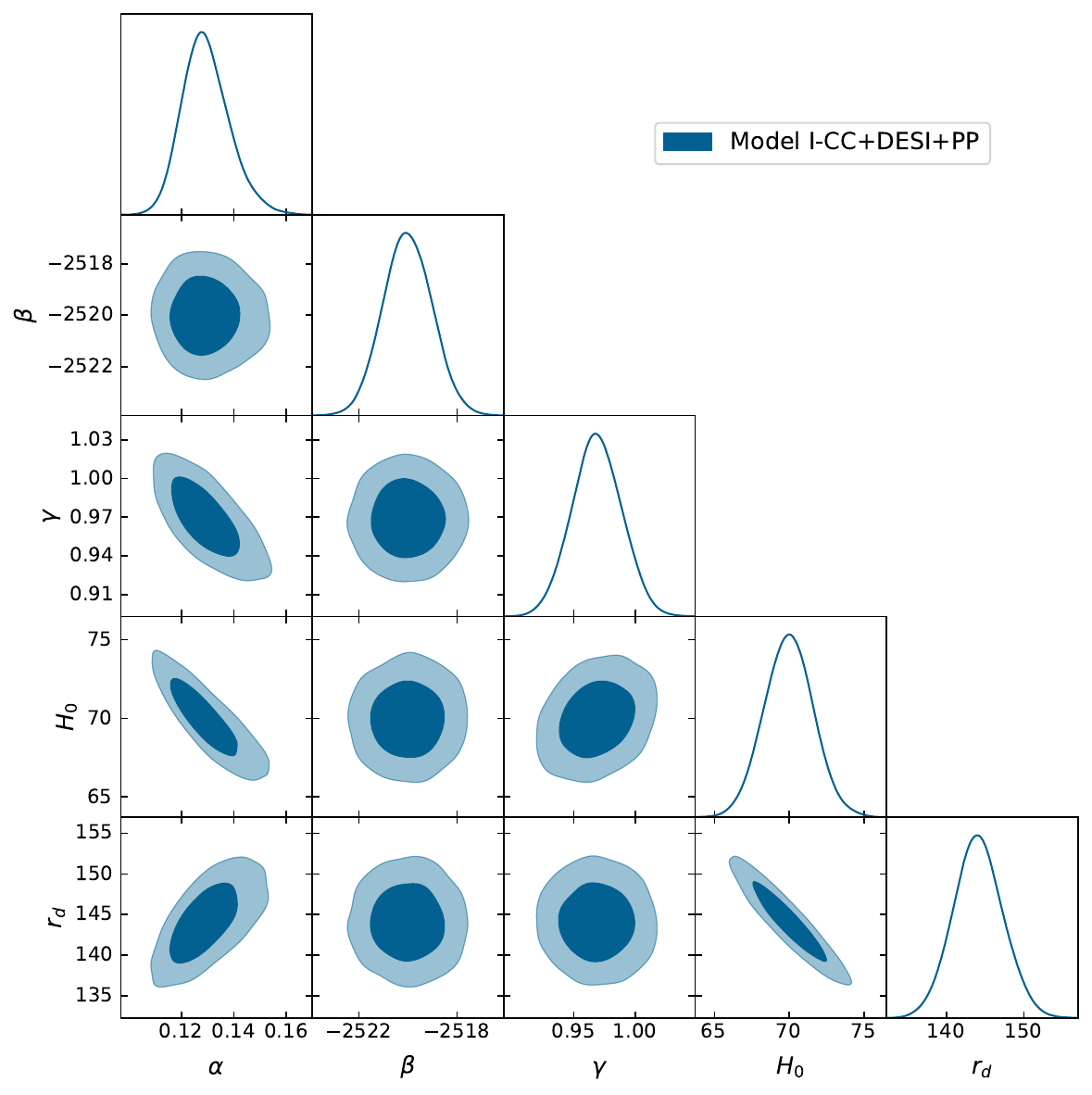}
\caption{\justifying Corner plot showing the posterior distributions and the \( 1\sigma \) and \( 2\sigma \) confidence contours for the parameter space \( ( \alpha, \beta, \gamma, H_{0},r_d) \) for Model I (\ref{sec: model_A}) within the \( f(Q, \mathcal{L}_m) \) gravity framework. }
\label{fig1}
\end{figure}

\begin{table}
\centering
\begin{tabular}{lccccccc}
\hline
\multicolumn{4}{c}{\textbf{68\% confidence level (CL) constraints}} \\ 
\hline \hline
Parameters &  Model I & Model II & Model III \\
 \hline \hline \\[-2ex] 
  $\alpha$  &  $0.1294^{+0.0076}_{-0.0099}$  & $0.1580^{+0.14}_{-0.077}$ & $0.10 \pm 0.058$\\ 

  $\beta$  &  $-2520 \pm 1.0$  & $-8830 \pm 0.049$ & $0.9880 \pm 0.072$ \\ 

 $\gamma$  & $0.969 \pm 0.020$  & $1.509 \pm 0.099$  & $0.836^{+0.087}_{-0.14}$\\ 

 $H_{0}$  &  $70.0 \pm 1.6$  & $70.1 \pm 1.6$ & $70.3 \pm 1.6$ \\ 

  $n$  &  $-$  & $-$ & $0.4010 \pm 0.01$ \\ 

 $r_d$  & $144.1 \pm 3.2$  & $143.8^{+3.0}_{-3.4}$ & $143.6 \pm 3.1$ \\ 

 $M_b$  & $-19.49^{+0.46}_{-0.35}$  & $-19.48 \pm 0.29$ & $-19.50\pm 0.29$\\ \hline \hline
\end{tabular}
\caption{\justifying Marginalized constraints and mean values with 68\% CL on the free model parameters.}
\label{tab:model_params}
\end{table}

\subsubsection*{Results for Model I}

The Fig.~\ref{fig1} displays the posterior distribution and parameter correlations for Model~I, derived from the combined CC+DESI+PP dataset. The diagonal panels show the marginalized one-dimensional posterior distributions for each parameter, while the off-diagonal panels present the corresponding two-dimensional joint distributions, with the inner and outer contours marking the \(68\%\) and \(95\%\) credible regions. The parameters are tightly constrained, with only mild degeneracies observed, most notably between \(H_{0}\) and \(r_{d}\). The compact contours indicate that the combined dataset provides strong constraints on \((\alpha, \beta, \gamma, H_{0}, r_{d})\), effectively breaking parameter degeneracies and yielding well-localized posterior peaks.

\begin{figure*}[htbp]
\centering

\begin{subfigure}[t]{0.34\textwidth}
    \centering
    \includegraphics[width=\textwidth]{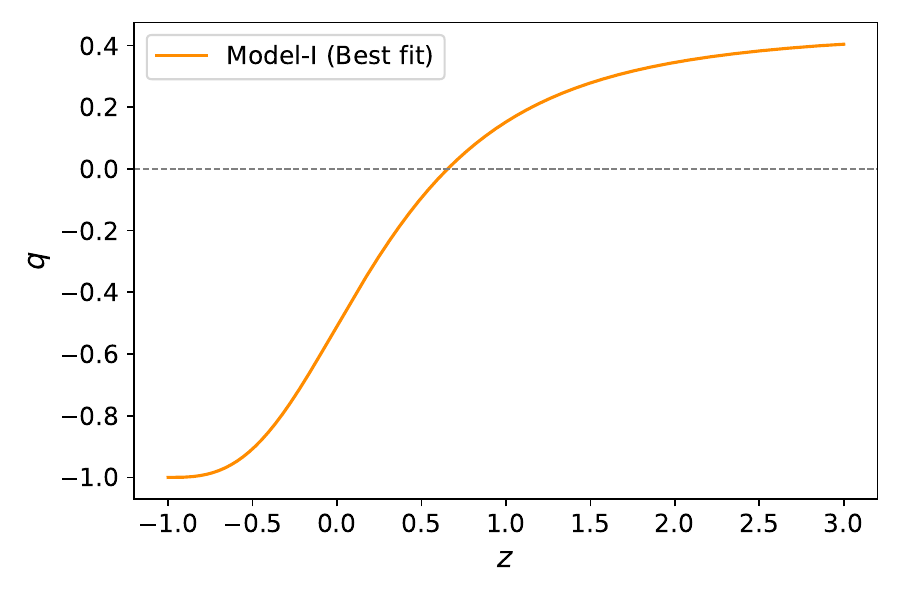}
   \caption{Deceleration parameter \(q(z)\)}
 \label{fig:qz-I}
\end{subfigure}%
\hspace{0.2cm}
\begin{subfigure}[t]{0.34\textwidth}
    \centering
    \includegraphics[width=\textwidth]{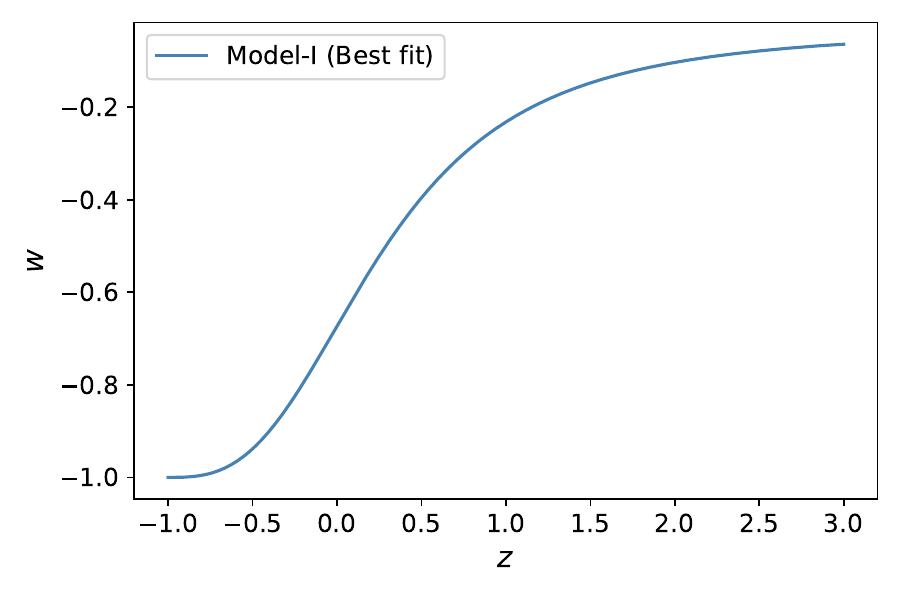}
 \caption{Equation-of-state parameter \(w(z)\)}
    \label{fig:wz-I}
\end{subfigure}%
\hspace{0.2cm}
\begin{subfigure}[t]{0.27\textwidth}
    \centering
    \includegraphics[width=\textwidth]{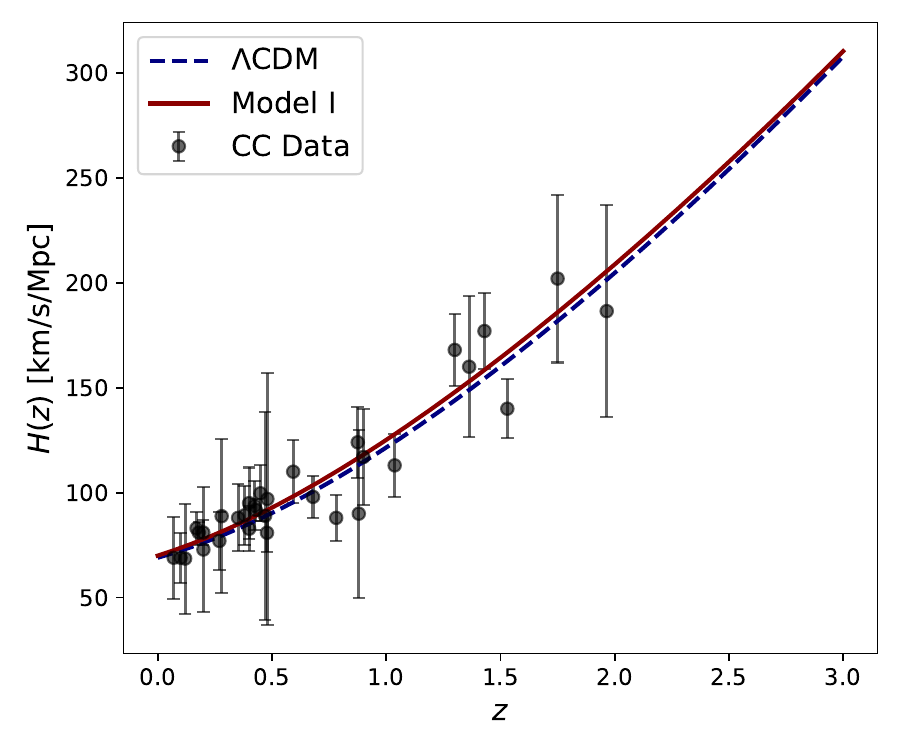}
\caption{Hubble parameter \( H(z)\)}
    \label{fig:Hz-I}
\end{subfigure}

\caption{\justifying
Redshift evolution of (a) the deceleration parameter \( q(z) \), (b) the equation-of-state parameter \( w(z) \), and (c) the Hubble parameter \( H(z) \), for Model I using the best-fit values of the model parameters. In (c), the dashed black curve represents the evolution of \( H(z) \) in the standard \( \Lambda \)CDM cosmological model with \( \Omega_m = 0.3 \) and \( H_0 = 69\)$Km/s/Mpc$, and observational Hubble data from cosmic chronometers are overlaid as black points with error bars. }
\label{fig_Cos_Model_I}
\end{figure*}

The Figs.~\ref{fig:qz-I} and \ref{fig:wz-I} illustrate the redshift evolution of the deceleration parameter \( q(z) \) and the dark energy equation-of-state parameter \( w(z) \), respectively, derived from MCMC analyses of the proposed cosmological model. In both plots, the solid curves represent the best-fit trajectories corresponding to the maximum-likelihood values of the model parameters. The evolution of \(q(z)\) exhibits a clear transition from a decelerating phase (\( q > 0 \)) at higher redshifts to an accelerating phase (\( q < 0 \)) at lower redshifts. This transition occurs at approximately \( z \approx 0.65 \), aligning well with the onset of late-time cosmic acceleration, with the present value estimated as \( q_{0} \approx -0.51 \)~\cite{mukherjee2021non,al2018observational,camarena2020local}.

Meanwhile, the behavior of \( w(z) \) shows a smooth evolution from \( w \approx 0 \) at early times, mimicking matter-like behavior, to \( w \approx -1 \) near the present epoch, aligning with a cosmological constant-like phase. The absence of phantom behavior (i.e., \( w < -1 \)) across the sampled trajectories and the consistency of both trends with observational constraints reinforce the robustness of the model in describing the Universe's expansion history.

The Fig.~\ref{fig:Hz-I} presents a comparison between the theoretical predictions of the Hubble parameter \( H(z) \) from two cosmological models and observational measurements obtained from cosmic chronometers (CC). The dashed blue curve corresponds to the standard \( \Lambda \)CDM model, while the solid red line represents the prediction from our proposed Model I. The black points with vertical error bars denote the observed \( H(z) \) values derived from CC data, which provide direct, model-independent measurements of the expansion rate at different redshifts. Both models exhibit excellent agreement with the data across the full redshift range \( 0 < z < 2 \), closely following the trend of the observed expansion history. The overlapping curves and the consistency with the data points reinforce the viability of Model I as a compelling alternative to the standard cosmological model.

\begin{table}[t]
		\begin{tabular}{l  c  c  r r r}
			\hline
	\textbf{Model} & \textbf{AIC} &\textbf{ BIC}  & $\boldsymbol{\chi^2 /\nu}$ & $ \boldsymbol{\Delta AIC}$ & $\boldsymbol{\Delta BIC}$\\
			\hline
			I & 1445.84 & 1478.22 & 0.8834 & 1.92 & 12.72 \\
			II& 1444.21& 1476.58 & 0.8824 & 0.29 &  11.08\\
			III & 1447.70 & 1485.47 & 0.8839 & 3.78 & 19.97 \\
			$\Lambda$CDM & 1443.92 & 1465.50 & 0.8836 & 0 & 0\\
			\hline
			\hline
		\end{tabular}
		\caption{The statistical comparison of the models.}
		\label{tab:stat}
	\end{table}

\subsection{Model II \texorpdfstring{$f= -\frac{Q}{2} + \alpha\, Q\,\mathcal{L}_m + \beta$ with $\mathcal{L}_m= p$}{}}\label{sec: model_II}

Now, we consider \( f(Q, \mathcal{L}_m) = -\frac{Q}{2} + \alpha\, Q\,\mathcal{L}_m + \beta \), where \( Q \) is the non-metricity scalar and \( \mathcal{L}_m \) is the matter Lagrangian. The first term recovers the STEGR, while the second introduces a non-minimal coupling between geometry and matter, analogous to extensions in \( f(R, \mathcal{L}_m) \) theories \cite{MontelongoGarcia:2010xd,Garcia:2010xb}. This coupling leads to modified dynamics and energy-momentum exchange, offering a mechanism for late-time acceleration. The constant \( \beta \) plays the role of an effective cosmological constant, allowing smooth recovery of \( \Lambda \)CDM in the appropriate limit.
The modified Friedmann equations in this model reduce to the following form:
\begin{align}
\label{b1}
3 H^{2} &= \frac{\beta - 3 H^2 (1 + 2\alpha \rho)}{4\alpha(\gamma - 1)\rho - 2} = \rho_{\mathrm{eff}}, \\
\label{b2}
2 \dot{H} + 3 H^{2} &= -p_{\mathrm{eff}},
\end{align}
where the effective pressure \( p_{\mathrm{eff}} \) is obtained from Eq.~\eqref{eq:friedmannmain} and is given by
\[
p_{\mathrm{eff}} = \frac{f}{4f_Q} + \left(1 - 2Q\frac{f_{QQ}}{f_Q} \right) \dot{H}.
\]

Due to the nonminimal coupling between the nonmetricity scalar \( Q \) and the matter Lagrangian \( \mathcal{L}_m \), the energy-momentum tensor is no longer conserved. This interaction leads to a modified cosmic evolution, distinct from standard GR. Given the nonlinearity of the differential equations, closed-form analytic solutions for \( H(z) \) are generally not feasible. Therefore, we solve the system numerically, imposing the initial condition \( H(0) = H_0 \), where \( H_0 \) is the present-day value.

\subsubsection*{Results for Model II}
\begin{figure}[]
\centering
\includegraphics[scale=0.42]{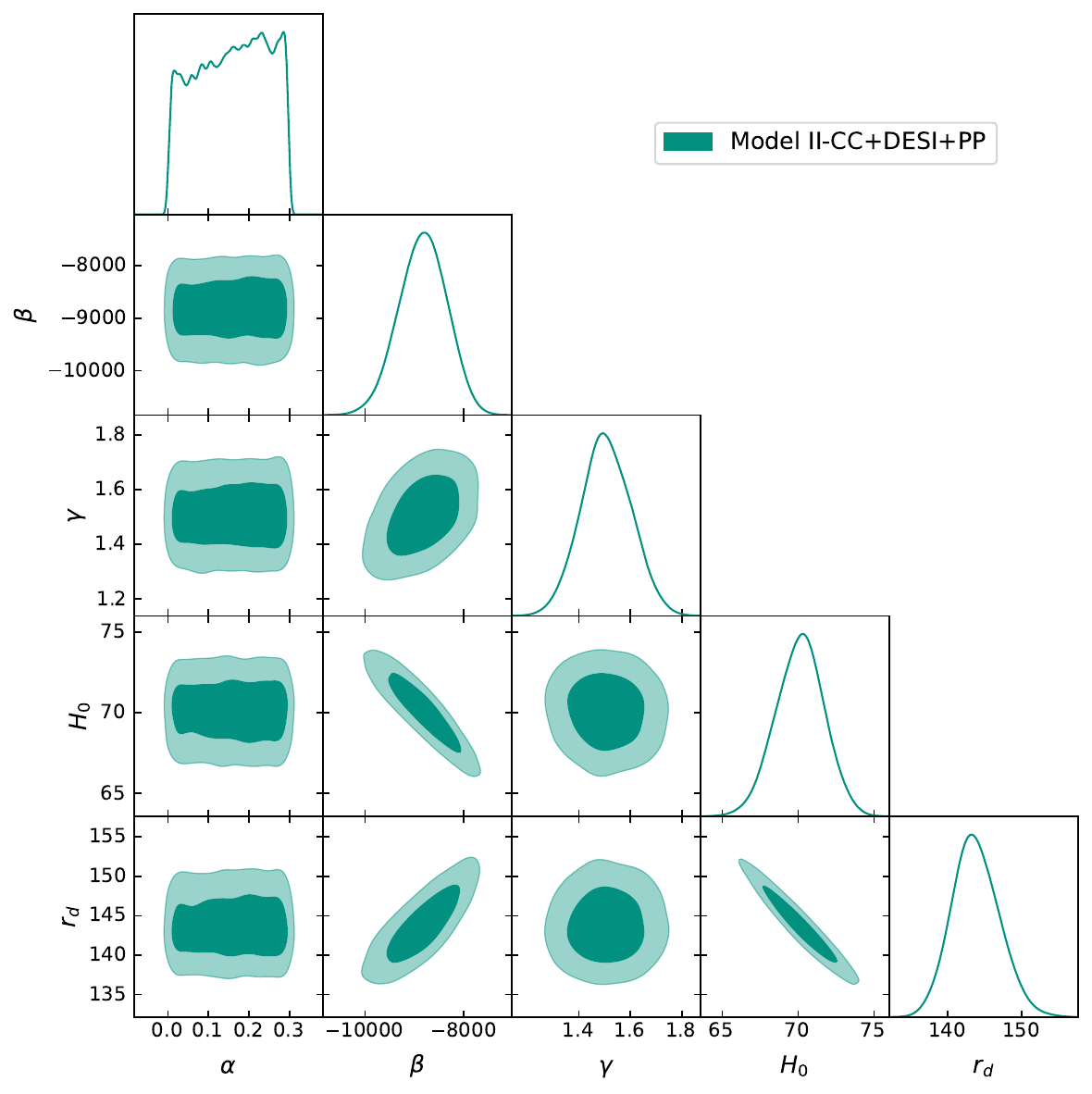}
\caption{\justifying Same as Fig. \ref{fig1}, but for Model II (\ref{sec: model_II}).}
\label{fig-TP2}
\end{figure}
The Fig.~\ref{fig-TP2} displays the one-dimensional marginalized posterior distributions along with the two-dimensional confidence contours at the \( 1\sigma \) and \( 2\sigma \) confidence levels. It presents the posterior distributions and parameter correlations for Model~II, obtained using the same CC+DESI+PP dataset. Compared to Model~I, Model~II exhibits broader posteriors for several parameters, reflecting the influence of its extended parameter space. In addition to the \(H_{0}\)–\(r_{d}\) correlation also seen in Model~I, a noticeable degeneracy appears between \(\beta\) and \(\gamma\). Despite these broader distributions, the constraints remain well-defined, indicating that the combined dataset still provides significant discriminatory power for this model.

\begin{figure*}[]
\centering

\begin{subfigure}[t]{0.34\textwidth}
    \centering
    \includegraphics[width=\textwidth]{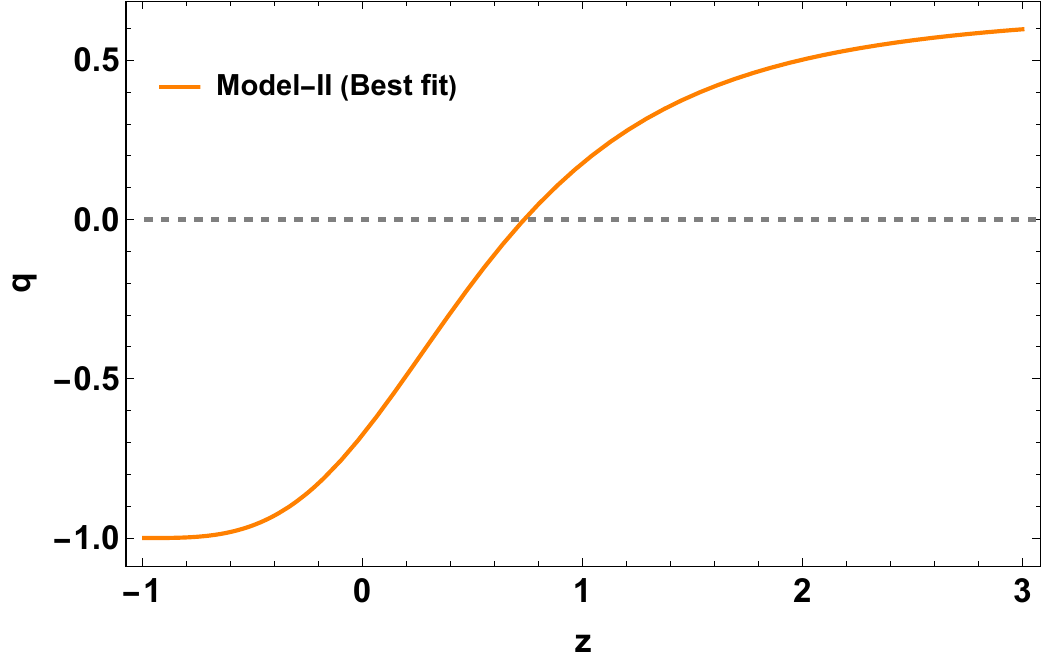}
   \caption{Deceleration parameter \(q(z)\)}
 \label{fig:qz-II}
\end{subfigure}%
\hspace{0.2cm}
\begin{subfigure}[t]{0.34\textwidth}
    \centering
    \includegraphics[width=\textwidth]{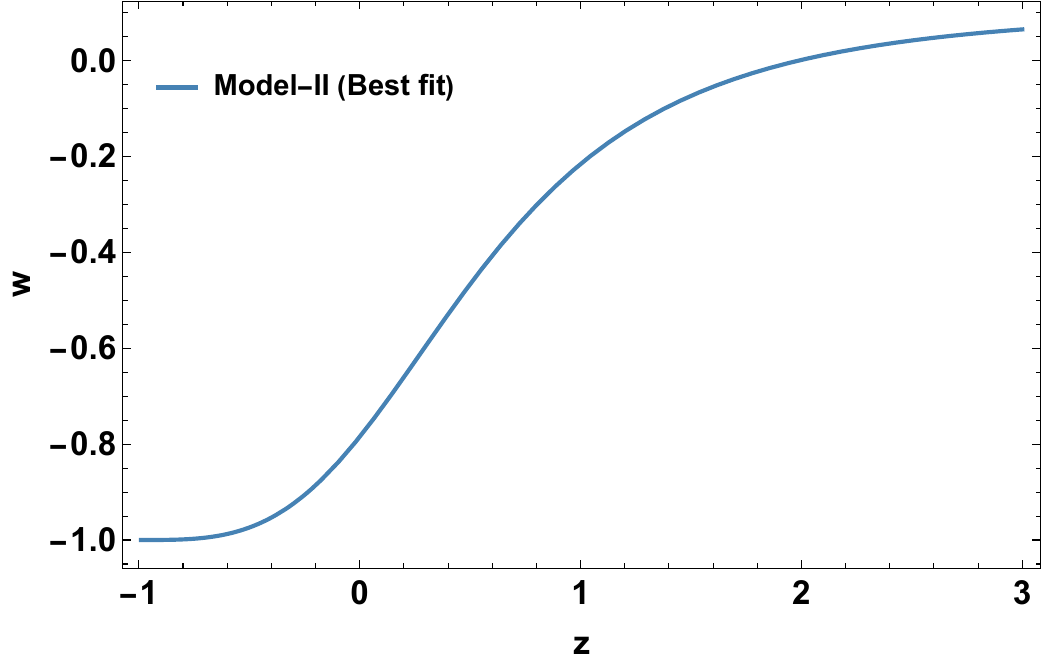}
 \caption{Equation-of-state parameter \(w(z)\)}
    \label{fig:wz-II}
\end{subfigure}%
\hspace{0.2cm}
\begin{subfigure}[t]{0.27\textwidth}
    \centering
    \includegraphics[width=\textwidth]{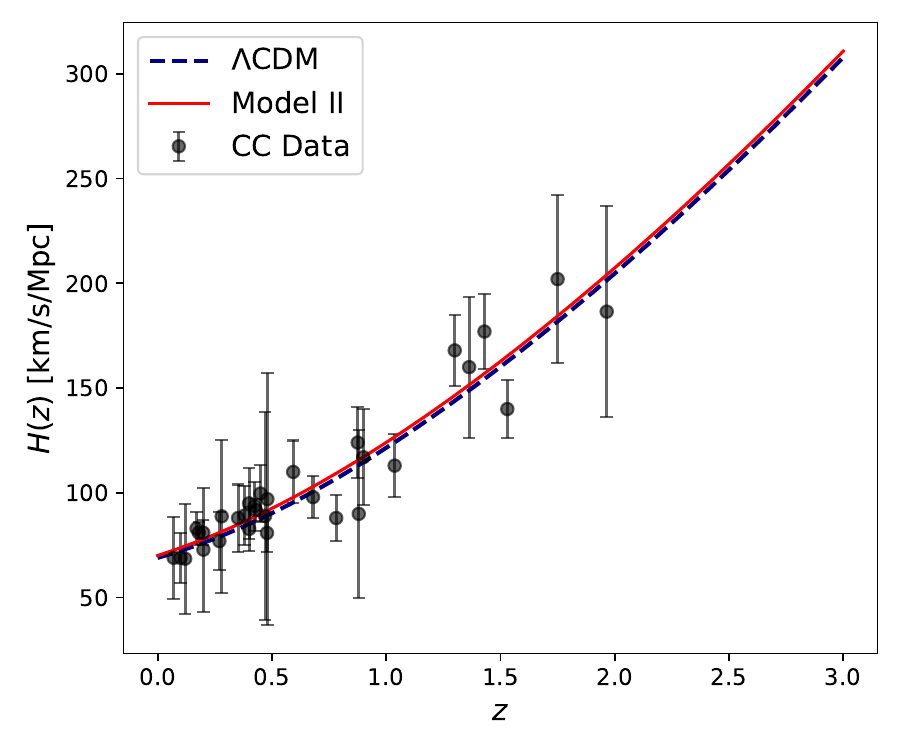}
\caption{Hubble parameter \(H(z)\)}
    \label{fig:Hz-II}
\end{subfigure}

\caption{\justifying Same as Fig. \ref{fig_Cos_Model_I}, but for Model II (\ref{sec: model_II}).}
\label{fig_Cos_Model_II}
\end{figure*}

The Fig.~\ref{fig_Cos_Model_II} presents the results for Model~II, plotted in the same manner as for Model~I. The deceleration parameter \( q(z) \) again exhibits a smooth transition from deceleration to acceleration, with a slightly later transition at \( z \approx 0.73 \) and a present value of \( q_{0} \approx -0.67 \). The effective equation-of-state parameter \( w(z) \) evolves from a quintessence-like regime (\( w \approx -1 \)) in the far future toward less negative values at higher redshifts. Notably, the present value differs between the two models, with \( w_{0} \approx -0.67 \) for Model~I and \( w_{0} \approx -0.78 \) for Model~II. In both cases, the predicted \( H(z) \) curves show excellent agreement with cosmic chronometer data and remain closely aligned with the \(\Lambda\)CDM predictions across the full redshift range.

\subsection{Model III \texorpdfstring{$f= -\frac{Q}{2} + \alpha\, Q^n \mathcal{L}_m + \beta$ with $\mathcal{L}_m= p$}{}}\label{sec: model_III}

Having examined the special case with \(n = 1\) in Model~II, we now turn to the more general scenario in which \(n\) is treated as a free parameter, leading to Model~III.
In this formulation, we consider a non-minimally coupled \( f(Q,\mathcal{L}_{m}) \) gravity model of the form \cite{MontelongoGarcia:2010xd,Garcia:2010xb}
\[
f(Q,\mathcal{L}_{m}) = -\frac{Q}{2} + \alpha\, Q^{n} \mathcal{L}_{m} + \beta,
\]
where \(n\) is treated as a free parameter alongside \(\alpha\) and \(\beta\). Unlike Model~II, where \(n\) was fixed to unity to yield a linear coupling between \(Q\) and \(\mathcal{L}_{m}\), Model~III allows \(n\) to vary and be constrained directly by observational data. This generalization increases the dimensionality of the parameter space and enables a more flexible assessment of the role of the non-minimal coupling exponent in shaping the cosmic evolution.
The modified Friedmann equations in this model reduce to the following form:
\begin{align}
\label{b1}
3 H^{2} &= \frac{\beta - 3 H^2 -\alpha (6H^2)^n \rho}{4\alpha \, n\, (\gamma - 1)\rho (6H^2)^{n-1} - 2} = \rho_{\mathrm{eff}}, \\
\label{b2}
2 \dot{H} + 3 H^{2} &= -p_{\mathrm{eff}},
\end{align}
where the effective pressure \( p_{\mathrm{eff}} \) is obtained from Eq.~\eqref{eq:friedmannmain} and is given by
\[
p_{\mathrm{eff}} = \frac{f}{4f_Q} + \left(1 - 2Q\frac{f_{QQ}}{f_Q} \right) \dot{H}.
\]
Given the nonlinear nature of the differential equations, closed-form analytic solutions for \( H(z) \) are generally unattainable. Consequently, we employ numerical integration, imposing the initial condition \( H(0) = H_{0} \), where \( H_{0} \) denotes the present-day Hubble parameter.

\subsubsection*{\textit{Results for Model III}}

\begin{figure}[]
\centering
\includegraphics[scale=0.45]{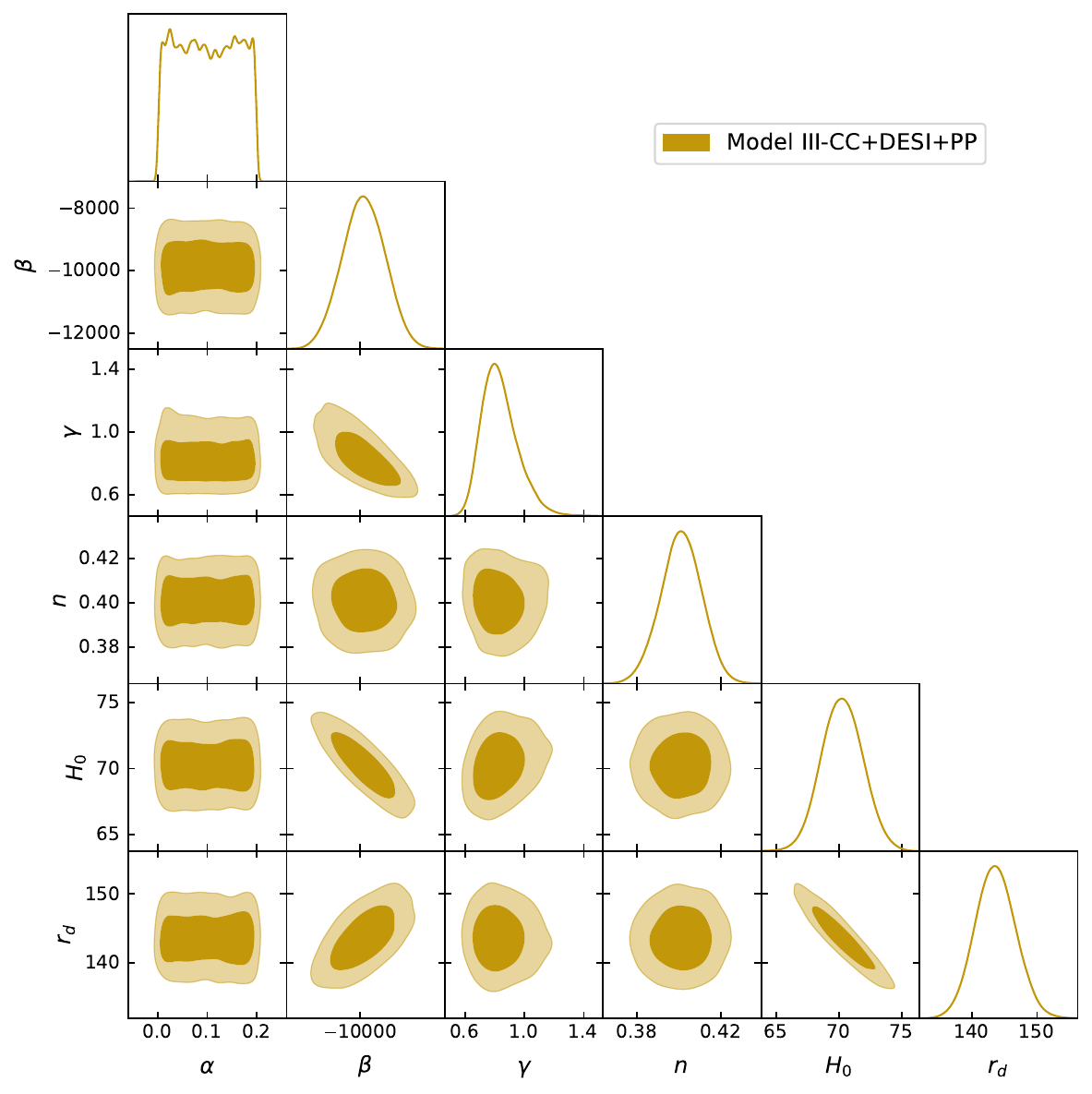}
\caption{\justifying Same as Fig. \ref{fig1}, but for the parameter space \( ( \alpha, \beta, \gamma, n, H_{0},r_d) \) for Model III (\ref{sec: model_III}).}
\label{fig-TP3}
\end{figure}

\begin{figure*}[]
\centering

\begin{subfigure}[t]{0.34\textwidth}
    \centering
    \includegraphics[width=\textwidth]{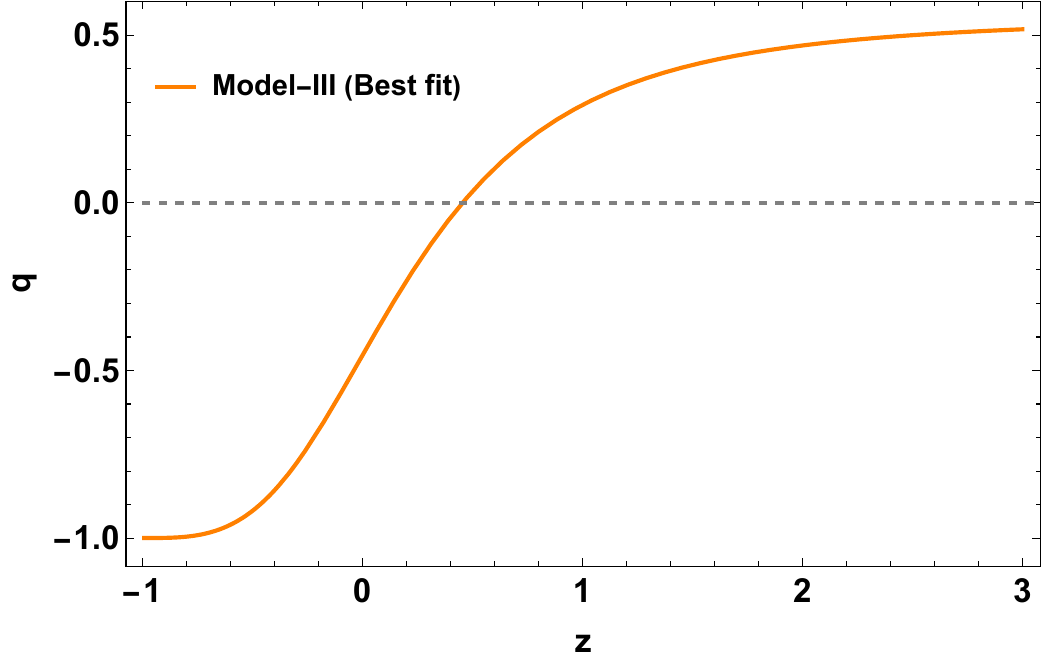}
   \caption{Deceleration parameter \(q(z)\)}
 \label{fig:qz-II}
\end{subfigure}%
\hspace{0.2cm}
\begin{subfigure}[t]{0.34\textwidth}
    \centering
    \includegraphics[width=\textwidth]{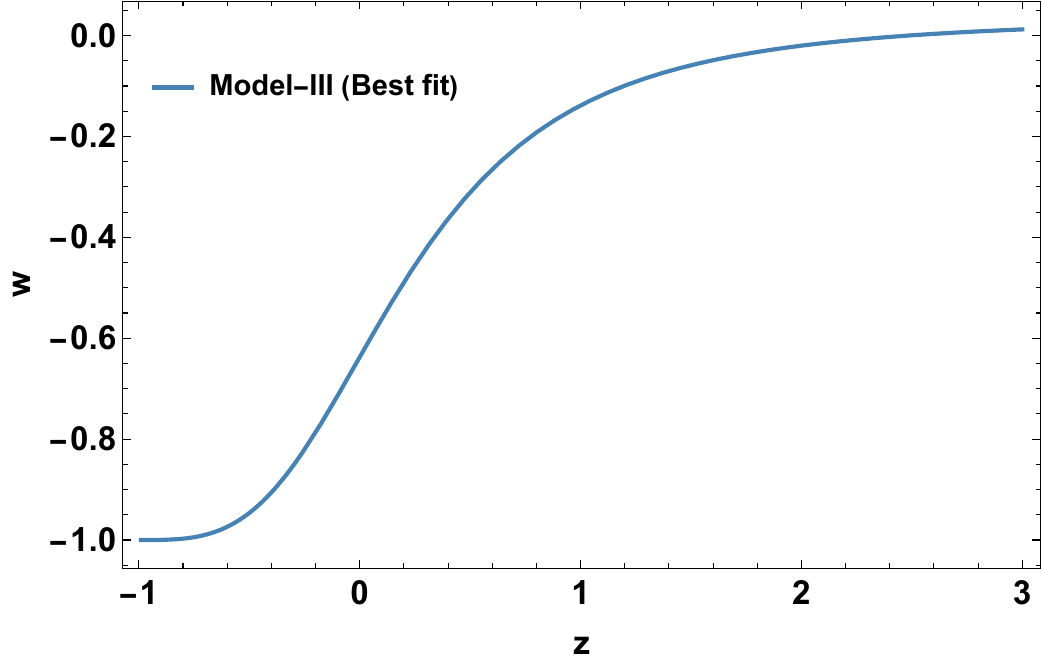}
 \caption{Equation-of-state parameter \(w(z)\)}
    \label{fig:wz-II}
\end{subfigure}%
\hspace{0.2cm}
\begin{subfigure}[t]{0.27\textwidth}
    \centering
    \includegraphics[width=\textwidth]{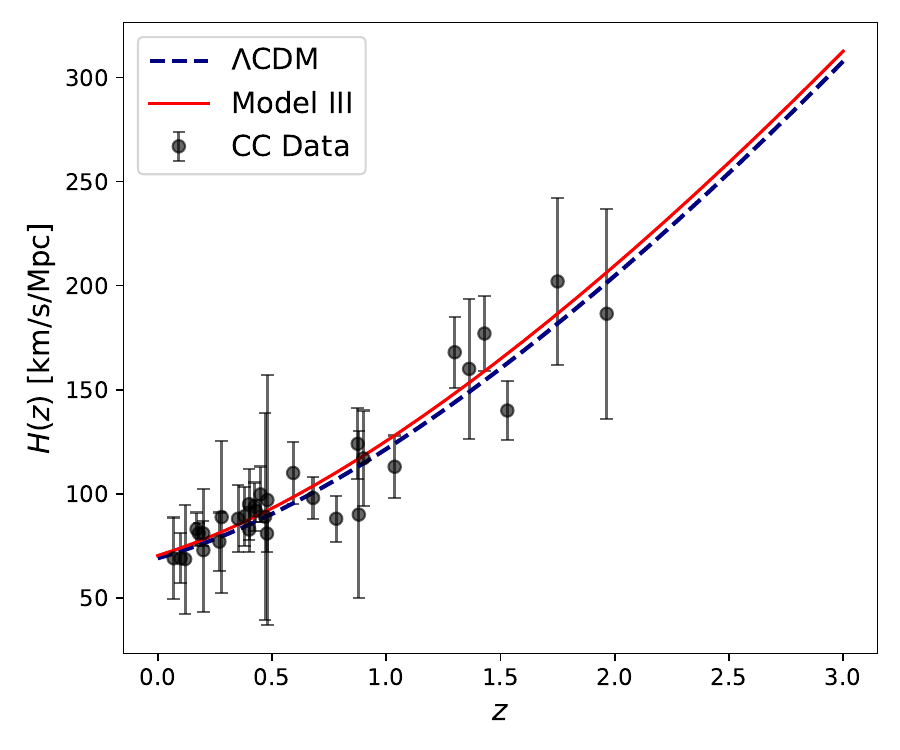}
\caption{Hubble parameter \(H(z)\)}
    \label{fig:Hz-II}
\end{subfigure}

\caption{\justifying Same as Fig. \ref{fig_Cos_Model_I}, but for Model III (\ref{sec: model_III}).}
\label{fig_Cos_Model_III}
\end{figure*}

The Fig.~\ref{fig-TP3} presents the posterior distributions for Model~III from the CC+DESI+PP dataset. Allowing \(n\) to vary enriches the parameter space and reveals meaningful correlations, particularly between \((\beta, \gamma)\), while maintaining well-constrained posterior regions. Fig.~\ref{fig_Cos_Model_III} summarises the best-fit cosmological evolution for Model~III. Compared with Model~II, freeing the exponent \(n\) shifts the deceleration–acceleration transition slightly later to \( z \approx 0.46 \) and yields a stronger present acceleration with \( q_{0} \approx -0.45 \). The corresponding equation-of-state parameter has \( w_{0} \approx -0.63 \), remaining close to quintessence-like behavior but showing a marginally less negative value than in Model~II. The \(H(z)\) profile maintains excellent consistency with CC data, with only minor deviations from the \(\Lambda\)CDM curve at intermediate redshifts, indicating that the additional freedom in \(n\) has a limited yet noticeable influence on the background expansion history.

\subsection{Statistical analysis}

The Table~\ref{tab:stat} presents the model selection statistics for all three proposed models alongside the \(\Lambda\)CDM baseline, evaluated using the combined CC+DESI+PP dataset. Among the extended models, Model~II achieves the lowest AIC and BIC relative to \(\Lambda\)CDM, indicating that it offers a slightly improved fit over Model~I and Model~III. All models yield comparable reduced chi-square values (\(\chi^{2}/\nu \approx 0.88\)), suggesting that they fit the data well. However, the larger \(\Delta\mathrm{BIC}\) values for the extended models, particularly Model~III, reflect the statistical penalty associated with their increased parameter space.

\begin{figure}[]
\centering
\includegraphics[scale=0.4]{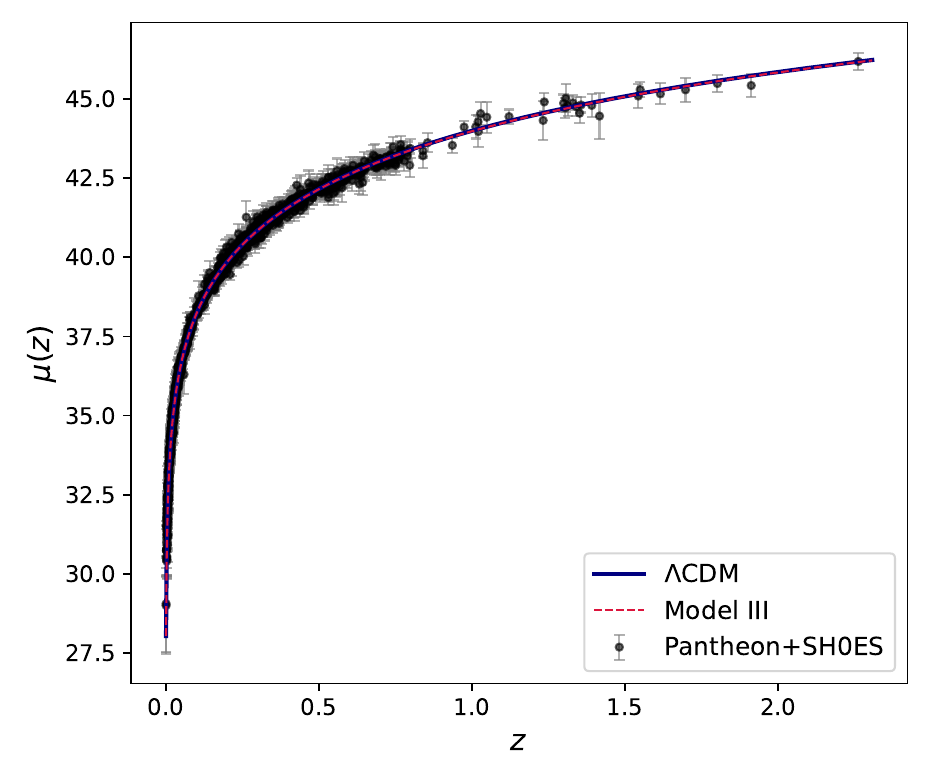} 
\caption{\justifying Behavior of the distance modulus \(\mu(z)\) as a function of redshift for Model~III. Since Models~I and II yield nearly indistinguishable curves on this scale, only Model~III is shown for clarity.}
\label{Error}
\end{figure}

Furthermore, Fig.~\ref{Error} shows the evolution of the distance modulus \(\mu(z)\) as a function of redshift for Model~III, compared with the \(\Lambda\)CDM prediction and the Pantheon+SH0ES supernova dataset. The Model~III curve closely follows the \(\Lambda\)CDM line across the full redshift range, reproducing the observed luminosity distance–redshift relation with high accuracy. The excellent agreement with the supernova data indicates that the additional freedom in \(n\) does not compromise consistency with low-redshift distance measurements. As the predictions for Models~I and II are nearly indistinguishable, we showed Model~III on this scale, for clarity.

\subsection{Correlation results}
\begin{figure}[]
\centering
\begin{subfigure}{0.35\textwidth}
    \centering
    \includegraphics[width=\linewidth]{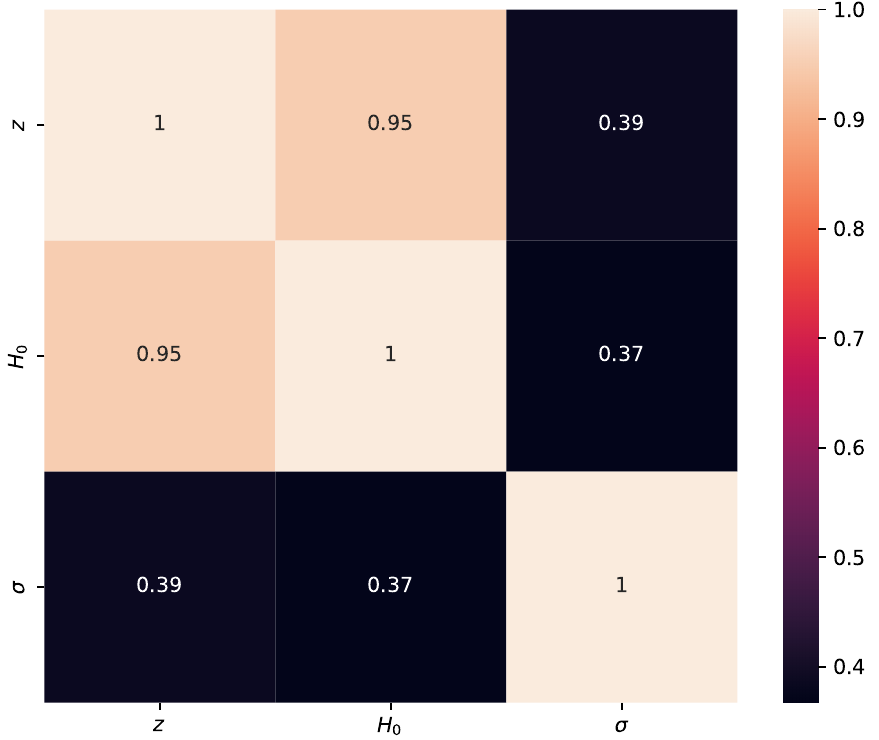}
    \subcaption{Hubble data}
    \label{fig:Hcor}
\end{subfigure}
\hspace{0.5cm}
\begin{subfigure}{0.35\textwidth}
    \centering
    \includegraphics[width=\linewidth]{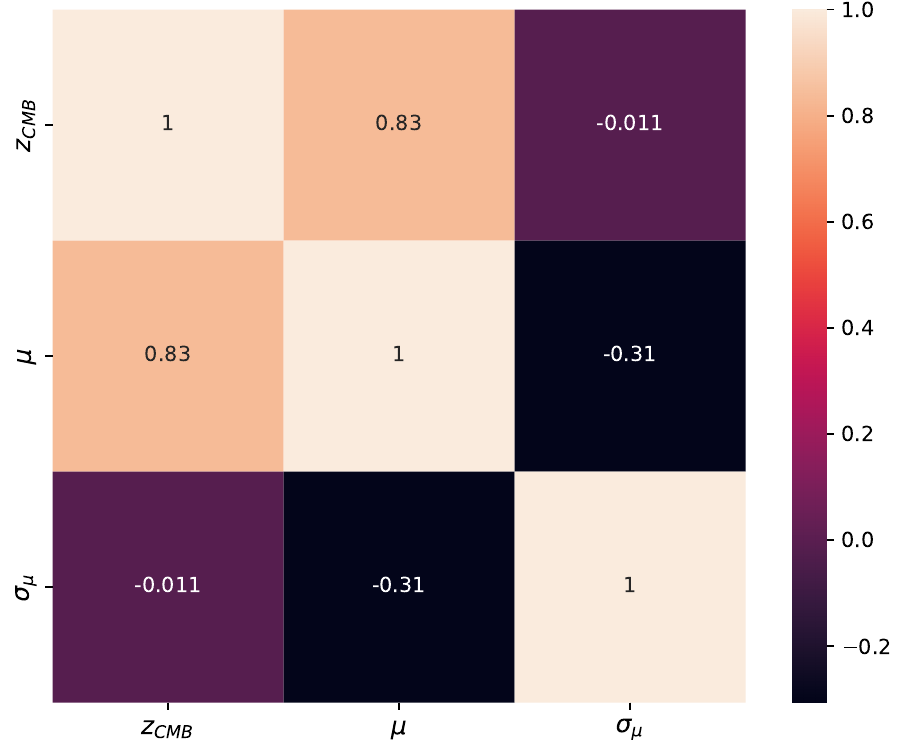}
    \subcaption{Pantheon+}
    \label{fig:Pcor}
\end{subfigure}
\caption{\justifying 
Correlation coefficients of different parameters from: (a) Hubble data and (b) Pantheon+.}
\label{fig:correlation_panels}
\end{figure}

We further investigated the Pearson correlation among key physical parameters within the datasets. From the results presented in Fig.~\ref{fig:correlation_panels}, we infer the following conclusions:
\begin{itemize}
    \item Fig. \ref{fig:Hcor} shows that the Hubble parameter $H_0$ exhibits a strong positive linear relationship with redshift $z$, as indicated by the correlation coefficient of $0.95$. The standard deviation $\sigma$ shows a moderate positive correlation with both $H_0$ (0.37) and $z$ (0.39).
    \item Fig. \ref{fig:Pcor} shows that $\mu$ has a strong positive correlation with $z_{CMB}$ (0.83) and moderate negative correlation with $\sigma_\mu$ (-0.31). The correlation coefficient of -0.011 indicates a very weak negative linear relationship between $\sigma_\mu$ and $z_{CMB}$.
\end{itemize}
For the DESI dataset, our analysis revealed no statistically significant correlations among its parameters; hence, we do not present its correlation panel here.

\subsection{Effective mass of scalar field particles}
As we have already mentioned, one of the important implications of $f\left(Q,\mathcal{L}_m\right)$ is related to the modification of the elementary particle equations at a fundamental level, leading to a generalization of the Klein-Gordon equation that explicitly contains nonmetricity and the matter Lagrangian. The presence of nonmetricity, as well as the non-minimal matter-geometry coupling leads to the modification of the particle mass $m_0^2$, which can be interpreted as an effective, cosmological parameters, and time-dependent mass.    

Using the results of section \ref{sec: cov}, we obtain the effective mass of a scalar field particle as given by Eq. \eqref{eq:meff} as
\begin{equation}
     m_{\mathrm{eff}}^2 = m_0^2 + m^{2(i)}_{Q\mathcal{L}_m}(t),
\end{equation}
where the superscript $(i)\in\{\text{I, II, III}\}$ labels the three model variants considered in this work. For $\xi=0$ the correction term vanishes, and the mass reduces to the constant bare value $m_0$. The term $m_{Q\mathcal{L}_m}^{2(i)}(t)$ gives a time-dependent correction to the mass of the particles. The model-dependent corrections take the explicit forms
\begin{equation}
\begin{aligned}
m^{2(\text{I})}_{Q\mathcal{L}_m}(t) =& \xi \left[3H^2(4-3\gamma)- \frac{3}{2}\left(\frac{\beta\gamma}{\alpha}\right)\right],\\
m^{2(\text{II})}_{Q\mathcal{L}_m}(t) =&\xi \left[3H^2 \left(\frac{3\gamma H^2 - \beta (7\gamma -4)}{3\gamma H^2 - \beta(\gamma -1)}\right) \right],\\
m^{2(\text{III})}_{Q\mathcal{L}_m}(t) =& \frac{\xi}{
  3H^2\big(1+(\gamma-1)n\big)-\beta(\gamma-1)n
}\bigg[\\
&\quad 6 \dot{H}\beta(\gamma-1)(n-1)n \\
&\qquad  + 9 H^4\big(4 -3\gamma + 4(\gamma-1)n \big)\\
  &\qquad \qquad  + 3 H^2\Big( 6 \dot{H}(\gamma-1)(n-1)n \\
  &\qquad \qquad \qquad  - \beta\big(3\gamma + 4(\gamma-1)n\big)\Big)\bigg].
\end{aligned}
\end{equation}
The correction term $m_{Q\mathcal{L}_m}^{2(i)}(t)$ is proportional to the coupling constant $\xi$ which directly determines whether the geometry–matter interaction increases ($\xi>0$) or decreases ($\xi<0$) the effective squared mass relative to the bare value. In Model I, for $\gamma =4/3$, the effective mass of the particle is a constant, $m_\mathrm{eff}^{2(\text{I})}=m_0^2-2\xi \beta /\alpha$, similar to the standard Klein-Gordon case, but with the mass still modified due to the presence of the geometry-matter coupling.  

However, statistical analysis of cosmological models suggests a value of $\gamma \approx 1$, which does not favor a constant effective scalar particle mass during cosmological evolution. For $\gamma=1$, $m_\mathrm{eff}^{2(\text{I})}=m_0^2 + 3\xi [H^2-(\beta /2\alpha)]$, and $m_\mathrm{eff}^{2(\text{II})}=m_\mathrm{eff}^{2(\text{III})}= m_0^2 + 3\xi (H^2-\beta)$. The effective mass is proportional to the square of the Hubble function $H^2$, and as such, it is a decreasing function of the cosmological time. In Model III, for $n=1$ it reduces to Model II. Model III exhibits the richest behavior, with terms proportional to both $H^4$ and $\dot{H}$, divided by a factor scaling as $H^2$ in the early universe. This structure amplifies the redshift dependence, making Model III generally more sensitive to background dynamics than Models I and II.

Since, as indicated by the statistical analysis and comparison with the observational cosmological data,  $\beta /\alpha <0$ and $\gamma \approx 1$ for all the models considered, it follows that the polarity of the correction term $m_{Q\mathcal{L}_m}^2(t)$ is solely determined by the coupling coefficient $\xi$ between the scalar field and geometry.

If we impose the condition of positivity of the effective mass, $m_\mathrm{eff}^2\geq 0$, we obtain an important constraint on the numerical value of the coupling parameter $\xi$, namely $\xi\geq -0.273$. Physically acceptable values of $\xi$ can thus be both positive and negative in the given range. On the other hand, a negative effective mass may appear in some condensed matter systems \cite{Yao_2008} called metamaterials. Generally, an object with a negative effective mass will have an acceleration opposite to the direction of the applied force.  However, in the following, we will discard this type of behavior as unphysical in a cosmological context.
  
\begin{figure*}[]
\centering
\includegraphics[scale=0.55]{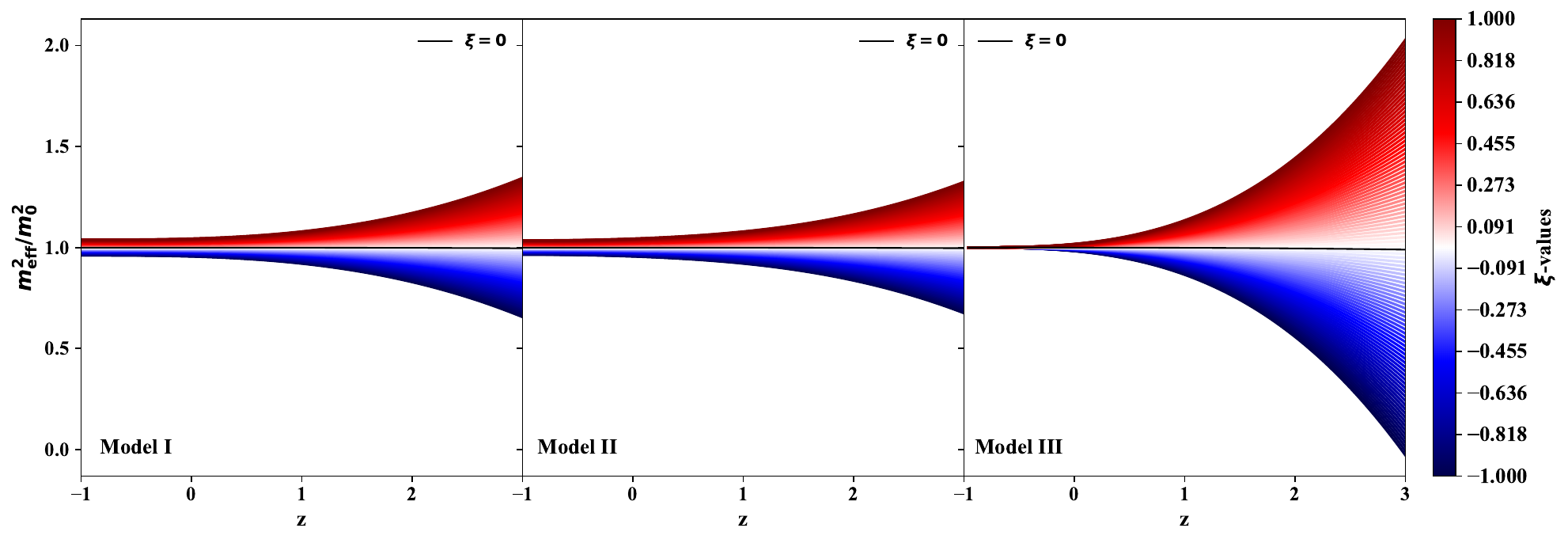}
\caption{\justifying The redshift variation of the ratio $m_\mathrm{eff}^2/m_0^2$ for different values of the parameter $\xi$. The left panel corresponds to Model I, the centre is Model II, while the right panel corresponds to Model III. The black solid line corresponds to the $\xi=0$ case, while the color gradient (blue to red) represents different values of $\xi$, ranging from -1 to 1. The data used in both models are from the combined CC+DESI+PP dataset, with the parameter values given in Table~\ref{tab:model_params}. We fixed the value of $m_{0}=6.583 \times 10^{-22}$ GeV to obtain the corresponding plots.}
\label{figmz}
\end{figure*}

The variation in the effective mass of the scalar particles is represented as a function of the redshift in Fig.~\ref{figmz}. The variation of $m_\mathrm{eff}^2$ depends significantly on the values of $\xi$. In Models I and II, for $\xi \in (-0.273,0)$, the ratio between the square of the effective mass and the remaining mass of the particle decreases to around 0.9. On the other hand, for $\xi >0$, the effective mass squared increases with the redshift, reaching a value of around 1.4 at a redshift of $z=3$. It is interesting to note that for all the Models, the variation of the effective mass is very similar. Models I and II show broadly similar behavior due to their common proportionality to $H^2$, whereas Model III displays significantly larger deviations from unity, reflecting its enhanced sensitivity to the background expansion rate. However, higher differences of the effective mass with respect to the standard rest mass are expected at higher redshifts, and this increase of the mass due to the geometry-matter coupling effects may have some significant implications on the dynamical behavior of particles in the very early stages of cosmological evolution. 

\section{Conclusions}\label{sec: discussion}

In this paper, we have investigated the theoretical aspects of the third geometric description of gravity, known as symmetric teleparallel gravity, or $f(Q)$ gravity. From a geometric and mathematical perspective, $f(Q)$ gravity uses the Weylian extension of Riemann geometry, where the fundamental metricity condition no longer holds. The violation of the metricity condition thus becomes the source of gravitational phenomena, with the non-metricity scalar $Q$ playing a similar role to that of the Ricci scalar in general relativity.
\par In the present study we have introduced a novel class of theories, representing an extension of $f(Q)$ gravity, where the non-metricity $Q$ is coupled non-minimally with the matter Lagrangian $\mathcal{L}_m$. Mathematically, our analysis was conducted within the framework of the metric-affine formalism. Our theory is constructed similarly to the $f(Q,T)$ theory, but with the trace of the matter energy-momentum tensor replaced by the matter Lagrangian. Similarly to the energy-momentum tensor trace-curvature couplings, in $f(Q,\mathcal{L}_m)$ theory, the coupling between $Q$ and $\mathcal{L}_m$ leads to the non-conservation of the matter energy-momentum tensor.
\par By applying a variational principle, we have derived the gravitational field equations for the $f(Q,\mathcal{L}_m)$ gravity theory. For particular choices of $f(Q,\mathcal{L}_m)$, it reduces to both $f(Q)$ and STEGR. This theory provides the freedom to explore different sets of coupling between $Q$ and $\mathcal{L}_m$, and thus the theory sheds light on the coupling mechanisms between the third, non-metric geometric description of gravity and matter, representing new avenues for further theoretical exploration. Consequently, the fundamental equations that describe the cosmological evolution in $f(Q,\mathcal{L}_m)$ gravity are expressed in terms of an effective energy density and pressure of a purely geometric origin. But they also depend on the ordinary matter-energy and pressure components of the energy-momentum tensor, as well as on the functions $f(Q,\mathcal{L}_m)$, $f_{Q}(Q,\mathcal{L}_m)$, and $f_{\mathcal{L}_m}(Q,\mathcal{L}_m)$.

Additionally, we have obtained the general relationship describing the non-conservation of the matter-energy-momentum tensor. The equation of motion of the particles reveals a notable departure from the geodesic motion for massive particles, specific to standard general relativity. An additional force emerges as a consequence of the coupling between $Q$ and $\mathcal{L}_m$. This coupling introduces a non-gravitational effect, leading to deviations from the paths followed in the standard geodesic motion and influencing the dynamical evolution of massive particles. The investigations presented may also contribute to a better understanding of the geometrical formulation of gravity theories, particularly regarding the aspects related to the geometry-matter coupling.

The standard tests of the gravitational field theories, including general relativity, are usually performed in vacuum. These standard tests involved the deflection of light by massive objects, the perihelion precession of the planets, geodesic motion, and Shapiro delay. In the case of the present $f\left(Q,\mathcal{L}_m\right)$ theory, in the vacuum limit $\mathcal{L}_m\rightarrow 0$, and $T_{\mu \nu}\rightarrow 0$, the field equations \eqref{eq:fieldequation} of the $f\left(Q,\mathcal{L}_m\right)$ theory reduce to the field equations of the $f(Q)$ theory \cite{heisenberg2024review}. Thus, all the vacuum tests of the two theories are the same. 

 In particular, black hole solutions and the propagation of gravitational waves coincide in the two theories. However, important differences are expected in the study of compact objects, like, for example, neutron stars. The structure of neutron stars in the $f(Q)$ theory was considered in \cite{Sokoliuk:2022bwi,Bhar:2023yrf}. In \cite{Bhar:2023yrf} it was shown that hybrid stars in the theory of $f(Q)$, satisfying a radial equation of state of the form $p_r=\alpha \rho-\beta$, where $\alpha$, $\beta$ are constants, can successfully model the observational characteristics of the Her X-1 star. Thus, $f(Q)$ gravity represents an attractive alternative in the description of compact objects. 

Similar investigations could also be performed within the framework of $f\left(Q,\mathcal{L}_m\right)$ gravity theory. The presence of the geometry-matter coupling leads naturally to an increase of the maximum allowable mass of compact stellar objects, and thus this coupling leads to a natural explanation of the high stellar masses of some neutron stars, which cannot be fully understood by using standard general relativity and the nuclear equations of state. The study of astrophysical objects could thus prove to be a testing ground for the present modified gravity theory, in which the observed masses of the neutron stars may lead to strong observational constraints on the parameters of the theory and on the functional form of $f\left(Q,\mathcal{L}_m\right)$.

Another possibility of testing the gravitational theory proposed in the present work is via the study of the geodesic deviation equation, which describes how objects moving under the influence of gravitational fields recede or approach one another. From an astrophysical point of view, the geodesic deviation equation has important applications in the study of the tidal forces, which have significant effects in the eccentric inspiralling neutron star binaries,  on the star formation in galaxies, due to the increase of the gas accretion rates as a result of the tidal perturbations induced by close stellar companions, and on the evolution of superradiant scalar-field states around spinning black holes \cite{yang2021geodesic}. 

As we have already seen from the analysis of the cosmological aspects of the analysis of $f\left(Q,\mathcal{L}_m\right)$ gravity theory, the curvature-matter coupling significantly modifies the nature of the gravitational interaction. A similar modification is also expected in relation to the tidal forces, as well as in the equation of motion in the Newtonian limit of the theory. Thus, a detailed comparison of the theoretical predictions of the $f\left(Q,\mathcal{L}_m\right)$ gravity theory related to the modifications of the tidal forces due to the presence of the geometry-matter coupling with the observational evidence obtained from the study of a large class of astrophysical phenomena could give some significant insights into the basic properties of the gravitational interaction, its geometric description, and constrain the effects of the nonmetricity in the Universe.

Hence, to obtain a consistent gravitational theory, one must consider its possibility of describing a large number of cosmological and astrophysical phenomena. Restricting the analysis of a given theory to only the cosmological (or astrophysical) framework may not provide enough evidence for its viability. Only testing the theory in various astrophysical/cosmological settings, which could be described in a consistent and non-contradictory way, with the same values of the coupling constants and of the functional form of the Lagrangian density of the theory, may give a full understanding of the theoretical and observational potential of a given theory.

An interesting effect of the matter-geometry coupling also appears when one considers the standard evolution equations of the elementary particles. 
In the present study, we have considered in detail the effects of $f\left(Q,\mathcal{L}_m\right)$ gravity on the Klein-Gordon equation, which describes the evolution of scalar particles, in the presence of the gravitational field whose effects are described by the Ricci scalar. By expressing the Ricci scalar with the help of the field equations, we have obtained a generalization of the Klein-Gordon equation that also explicitly includes, beyond the effects of the nonmetricity, the effects of the geometry-matter coupling, described by the matter Lagrangian, the trace of the matter energy-momentum tensor, as well as the derivatives of the Lagrangian density $f\left(Q,\mathcal{L}_m\right)$ with respect to $\mathcal{L}_m$. All these extra effects can be combined in a single term that gives an effective contribution to the rest mass of the particle $m_0$. Hence, the modified gravity effects generate an effective mass, which may have important implications on the scalar particle evolution in the early Universe. For both cosmological models considered, the effective mass is proportional to $H^2$, and, for the obtained values of the optimal model parameters, the sign of the effective mass is determined by the coupling parameter $\xi$ between the geometry and the scalar field. The condition of the positivity of the effective mass allows us to obtain some constraints on the value of $\xi$. The redshift variation of $m_\mathrm{eff}^2$ also depends on the sign and numerical values of $\xi$, and thus the effective mass can increase or decrease during cosmological evolution. This variation of the effective mass resulting from the coupling between matter and geometry, as well as from the presence of nonmetricity,  could have potentially important implications for the behavior of the scalar fields in both early and late Universe, in phenomena like inflation, reheating, Big Bang nucleosynthesis, or the recent accelerated expansion.            

For the description of the dynamics of the Universe, we have adopted the homogeneous and isotropic FLRW-type metric, describing the cosmological evolution in a flat geometry. In this study, we have examined three specific classes of cosmological models by adopting some functional forms of $f(Q,\mathcal{L}_m)$. For the first case, we have assumed the simple additive Lagrangian, $f(Q,\mathcal{L}_m)=-\alpha Q+2\mathcal{L}_m+\beta$. For this model, we have obtained a wide range of cosmological scenarios and evolution corresponding to the specific numerical values of the model parameters. These scenarios may include cosmological evolution that describes both the decelerating and the accelerating expansion phases of the Universe and de Sitter-type dynamics at late times. The model $f(Q,\mathcal{L}_m)=-\alpha Q+2\mathcal{L}_m+\beta$ can provide an effective description of cosmological data up to redshifts of around $z \approx 1$. Specifically, in this model, the Universe undergoes a rapid transition from a decelerating phase, characterized by a positive value of $q$, to an accelerating state where $q<0$. This transition can result, in its final stages, in a de Sitter-type expansion. 

The second and third model with $f(Q,\mathcal{L}_m)=-\frac{Q}{2} + \alpha\, Q^n \mathcal{L}_m+\beta$ for $n=1$ and free $n$ also evolves from a decelerating state to an accelerating state. Both models exhibit excellent consistency with cosmic chronometer measurements and remain closely aligned with $\Lambda$CDM predictions across the explored redshift range. In Model~II, the Universe transitions from deceleration to acceleration at $z \approx 0.73$, with a present value of $q_{0} \approx -0.67$ and an effective equation-of-state parameter $w_{0} \approx -0.78$, indicative of a strong quintessence-like behavior. In Model~III, freeing the exponent $n$ expands the parameter space, revealing meaningful correlations, particularly between $(\beta, \gamma)$, while maintaining tight posterior constraints. This added flexibility shifts the deceleration–acceleration transition to a later epoch at $z \approx 0.46$ and yields a slightly weaker present acceleration ($q_{0} \approx -0.45$) with $w_{0} \approx -0.63$, still consistent with a quintessence-like regime. While the influence of $n$ on the overall background expansion is modest, it introduces measurable differences in late-time dynamics, suggesting that higher-order model extensions could further refine constraints on cosmic acceleration.

Another potential application of $f(Q,\mathcal{L}_m)$ theory would be to consider inflation in the presence of scalar fields, which might offer a completely new perspective on the geometrical, gravitational, and cosmological processes that significantly influenced the early dynamics of the Universe. Consequently, the predictions of the present model could lead to major differences compared to those of standard general relativity or its extensions that ignore the role of matter. These differences could impact several current areas of interest, such as cosmology, gravitational collapse, and the generation of gravitational waves. In conclusion, in the present investigation, we have introduced a new version of the symmetric teleparallel theory and demonstrated its theoretical consistency. This approach also motivates and encourages the exploration of further extensions within the $f(Q,\mathcal{L}_m)$ family of theories.

\appendix
\setcounter{equation}{0}
\renewcommand{\theequation}{A\arabic{equation}}
\section{Derivation of the Friedmann equations}\label{sec:Friedmann equations}
The metric tensor components are given by $g_{\mu\nu}={\rm diag}\left(-1,a^2,a^2,a^2\right)$,  $g^{\mu\nu}={\rm diag}\left(-1,a^{-2},a^{-2},a^{-2}\right)$, and its determinant $\sqrt{-g}=a^3$. For the nonmetricity tensor, have the following nonzero terms,
\begin{gather}
    Q_{011}=Q_{022}=Q_{033}=2a\dot{a},\\
    Q_{0}^{\;\;11}=Q_{0}^{\;\;22}=Q_{0}^{\;\;33}=\frac{2\dot{a}}{a^3},\\
    Q^{01}_{\;\;\;1}=Q^{02}_{\;\;\;2}=Q^{03}_{\;\;\;3}= -\frac{2\dot{a}}{a},\\
    L^0_{\;\;11}=L^0_{\;\;22}=L^0_{\;\;33}=-a\dot{a},\\
    L^1_{\;\;01}=L^1_{\;\;10}=L^2_{\;\;02}=L^2_{\;\;20}=L^3_{\;\;03}= L^3_{\;\;30}= -\frac{\dot{a}}{a},\\
    P^0_{\;\;11}=P^0_{\;\;22}=P^0_{\;\;33}=-a\dot{a},\\
    P^{011}=P^{022}=P^{033}=-\frac{\dot{a}}{a^3},\\
    P_{011}=P_{022}=P_{033}=a\dot{a},\\    P^1_{\;\;01}=P^1_{\;\;10}=P^2_{\;\;02}=P^2_{\;\;20}=P^3_{\;\;03}=P^3_{\;\;30}=-\frac{\dot{a}}{4a},\\
    P_{110}=P_{101}=P_{220}=P_{202}=P_{330}=P_{303}=-\frac{a\dot{a}}{4},\\
    P^{110}=P^{101}=P^{220}=P^{202}=P^{330}=P^{303}=-\frac{\dot{a}}{4a^3}.
\end{gather}
The non-metricity scalar $Q$ is calculated using Eq.\eqref{eq:non-metricityscalar} as
\begin{equation}
    Q=-(Q_{011}P^{011}+Q_{022}P^{022}+Q_{033}P^{033}).
\end{equation}
We thus obtain $Q=6H^2$, where $H=\dot{a}/a$. 

The energy-momentum tensor $T_{\mu\nu}$ for a perfect fluid has the components 
\begin{equation}
T_{\mu\nu}={\rm diag}(\rho,pa^2,pa^2,pa^2). 
\end{equation}

Evaluating the field equation \eqref{eq:fieldequation} for the tt-component

\begin{equation}
    \begin{aligned}
    \frac{2}{a^3}\nabla_\alpha(f_Q\sqrt{-g}P^\alpha_{\;\;00})+f_Q(P_{0\alpha\beta}Q_0^{\;\;\alpha\beta}-2Q^{\alpha\beta}_{\;\;\;0}P_{\alpha\beta 0})\\
    +\frac{1}{2}f g_{00}=
    \frac{1}{2}f_{\mathcal{L}_m}(g_{00}\mathcal{L}_m-T_{00}),
    \end{aligned}
\end{equation}

\begin{equation}
    \begin{aligned}
f_Q(P_{011}Q_0^{\;\;11}+P_{022}Q_0^{\;\;22}+P_{033}Q_0^{\;\;33})-\frac{1}{2}f=-\frac{1}{2}f_{\mathcal{L}_m}(\rho+\mathcal{L}_m),
    \end{aligned}
\end{equation}
gives the first generalized Friedmann equation
\begin{equation}
    3H^2=\frac{1}{4f_Q}\bigr[ f - f_{\mathcal{L}_m}(\rho + \mathcal{L}_m) \bigl].
\end{equation}

By evaluating the field equation \eqref{eq:fieldequation} for the xx-component

\begin{equation}
    \begin{aligned}
    \frac{2}{a^3}\nabla_\alpha(f_Q\sqrt{-g}P^\alpha_{\;\;11})+f_Q(P_{1\alpha\beta}Q_1^{\;\;\alpha\beta}-2Q^{\alpha\beta}_{\;\;\;\;1}P_{\alpha\beta 1})\\
    +\frac{1}{2}f g_{11}=\frac{1}{2}f_{\mathcal{L}_m}(g_{11}\mathcal{L}_m-T_{11}),
    \end{aligned}
\end{equation}

\begin{equation}
    \begin{aligned}
    \frac{2}{a^3}\frac{\partial}{\partial t}(f_Qa^3(-a\dot{a}))-2f_Q\bigr(\frac{2\dot{a}}{a}\bigl)(a\dot{a})
    +\frac{a^2}{2}f=\frac{a^2}{2}f_{\mathcal{L}_m}(\mathcal{L}_m-p),
    \end{aligned}
\end{equation}
leads to the second generalized Friedmann equation
\begin{equation}
    \dot{H} + 3H^2 + \frac{\dot{f_Q}}{f_Q}H=\frac{1}{4f_Q}\bigr[ f + f_{\mathcal{L}_m}(p - \mathcal{L}_m) \bigl].
\end{equation}



\section*{Data Availability}
Data sharing is not applicable to this article as no datasets were generated or analyzed during the current study.

\section*{Acknowledgments} AH expresses its gratitude to Mr. Sayantan Ghosh (BITS-Pilani, Hyderabad) for insightful discussions on modified gravity, and extends thanks to Dr. Subhadip Sau (Jhargram Raj College) and Dr. Massimo Rossi (INAF/OAS Bologna) for valuable conversations regarding the Klein-Gordon equation.  SA acknowledges the Japan Society for the Promotion of Science (JSPS) for providing a postdoctoral fellowship during 2024-2026 (JSPS ID No.: P24318). This work of SA is supported by the JSPS KAKENHI grant (Number: 24KF0229). PKS acknowledges Anusandhan National Research Foundation (ANRF),  Department of Science and Technology, Government of India for financial support to carry out Research project No.: CRG/2022/001847 and IUCAA, Pune, India, for providing support through the visiting Associateship program. We would like to thank the anonymous reviewer for comments and suggestions that helped us to significantly improve our work.





\bibliography{references}

\end{document}